\documentclass[aps,prb,twocolumn,superscriptaddress,longbibliography]{revtex4-2}

\usepackage[colorlinks=true,citecolor=blue]{hyperref}
\usepackage{graphicx}
\usepackage{amsmath}
\usepackage{amssymb}
\usepackage{appendix}

\usepackage{xcolor}
\usepackage{ulem}
\usepackage{adjustbox}
\usepackage{braket}

\begin{document}



\title{Machine learning the Kondo entanglement cloud from local measurements}

\author{Faluke Aikebaier}
\affiliation{Department of Applied Physics, Aalto University, 00076, Espoo, Finland}
\affiliation{Computational Physics Laboratory, Physics Unit, Faculty of Engineering and Natural Sciences, Tampere University, FI-33014 Tampere, Finland}
\affiliation{Helsinki Institute of Physics P.O. Box 64, FI-00014, Finland}

\author{Teemu Ojanen}
\affiliation{Computational Physics Laboratory, Physics Unit, Faculty of Engineering and Natural Sciences, Tampere University, FI-33014 Tampere, Finland}
\affiliation{Helsinki Institute of Physics P.O. Box 64, FI-00014, Finland}

\author{Jose L. Lado}
\affiliation{Department of Applied Physics, Aalto University, 00076, Espoo, Finland}

\begin{abstract}
A quantum coherent screening cloud around a magnetic impurity in metallic systems is the hallmark of the antiferromagnetic Kondo effect. Despite the central role of the Kondo effect in quantum materials, the structure of quantum correlations of the screening cloud has defied direct observations. In this work, we introduce a machine-learning algorithm that allows to spatially map the entangled electronic modes in the vicinity of the impurity site from experimentally accessible data. 
We demonstrate that local correlators allow reconstructing the local many-body correlation entropy in real-space in a double Kondo system with overlapping entanglement clouds. Our machine learning methodology allows 
bypassing the typical requirement of measuring long-range non-local correlators with conventional methods.
We show that our machine learning algorithm is transferable between different Kondo system sizes, and we show its robustness in the presence of noisy correlators. Our work establishes the potential of machine learning methods to map many-body entanglement from real-space measurements.
\end{abstract}

\maketitle

\section{Introduction}

Strongly interacting quantum many-body systems exhibit a wealth of intricate physical phenomena. 
Quantum impurity problems, and in particular the Kondo problem~\cite{deHaas1934,Kondo1964,RevModPhys.55.331}, 
play a crucial role in capturing properties of the localized interactions within a larger quantum system~\cite{Hewson1993,PhysRevLett.85.2557,PhysRevLett.88.096804,PhysRevLett.85.2557}.
Such systems provide a paradigmatic framework for understanding the correlation effects and related entanglement features in many-body systems~\cite{Cho2006,Affleck2009,Wirth2016,RevModPhys.69.809}. 
A hallmark feature of the Kondo effect is the formation of a dynamic cloud of conduction electrons, or "the Kondo screening cloud", surrounding the impurity. The Kondo cloud, which  plays a crucial role in understanding the Kondo problem~\cite{Mukherjee2022}, leads
to electron entanglement at mesoscopic scales~\cite{Barzykin1996,Affleck2010}. 
Recent experiments have directly confirmed the existence of the Kondo screening cloud~\cite{VBorzenets2020}, however, the detailed structure of the quantum many-body correlations remains elusive. 
Correlation effects are essential for understanding the emergence of the Kondo effect and the subsequent formation of the Kondo screening cloud~\cite{Gubernatis1987,Barzykin1996,Barzykin1998,Borda2007,Holzner2009}, motivating the development of more powerful strategies
to imaging the Kondo entanglement cloud.

\begin{figure}[t!]
\includegraphics[width=0.8\linewidth]{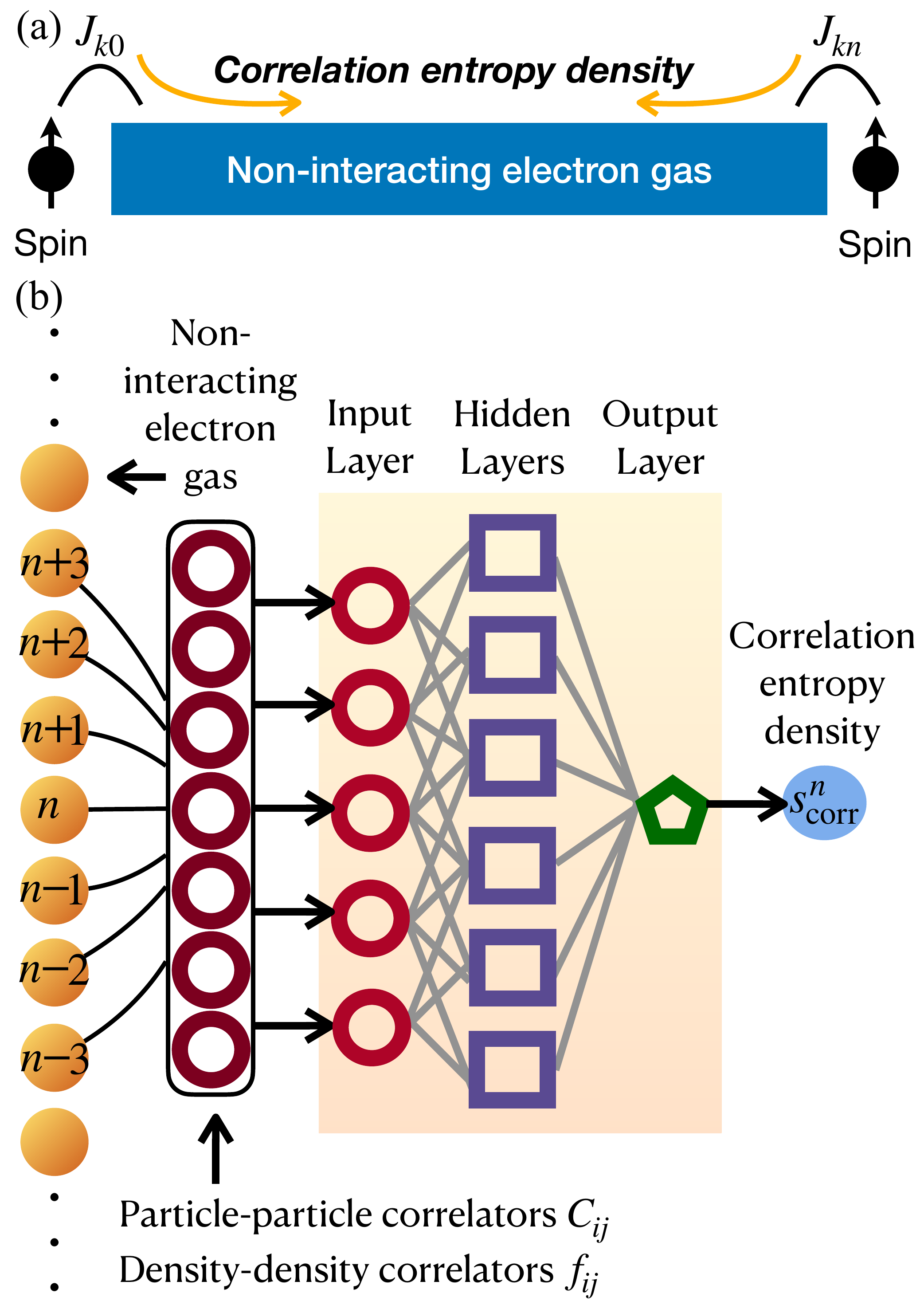}%
\caption{\label{fig:setup} (a) Schematic of the model. Two interacting spins on the two sides of the non-interacting chain induces many-body correlation via Kondo coupling. (b) Schematic of the workflow of the neural-network model, taking as input spatially resolved local
correlators, and providing as output the spatial profile of the correlation entropy density.}
\end{figure}

Entanglement properties of quantum materials are remarkably challenging to extract in experiments. 
From a theory perspective,
correlations in electronic systems can be quantified by means of the von Neumann entropy obtained from one-particle density matrix, known as the correlation entropy~\cite{He2014,Lu2014,Fishman2015,Esquivel1996,Gersdorf1997,Ziesche1997},
a quantity that vanishes for any non-interacting electronic system. 
Experimental measurement of the correlation entropy is greatly challenging
as it requires knowledge of all correlators in the whole system\cite{Huang2006,BenavidesRiveros2017,Ferreira2022}. 
Machine learning methodologies algorithm offer a potential alternative strategy
for extracting the correlation entropy from a reduced set of measurements~\cite{Aikebaier2023}. 
Machine learning methods have been demonstrated to be highly successful in extracting Hamiltonians
from experimental data\cite{PhysRevA.80.022333,Wang2017,Hincks2018,PhysRevApplied.20.024054,Khosravian2024,PhysRevA.105.023302,PhysRevResearch.3.023246,PhysRevResearch.1.033092,PhysRevResearch.4.033223,PhysRevLett.102.187203,PhysRevApplied.20.044081}, and
and for automatic tuning of quantum systems
without human intervention\cite{Schmale2022,PhysRevA.102.042604,Torlai2018,Quek2021,Palmieri2020,PhysRevLett.127.140502,PhysRevA.106.012409,Cha2021}. 
However, its potential for extracting local entanglement properties in homogeneous
single impurity Kondo problems remains unexplored. 

In this work, we develop a machine-learning assisted algorithm, which employs local measurements near a Kondo
magnetic impurity to predict the spatial entanglement in real-space.
We demonstrate that a supervised machine learning approach allows predicting the
spatially varying correlation entropy density solely from local correlations. 
This method enables one to extract the spatial structure of the quantum correlations in the Kondo screening cloud,
as well as the overlap between two Kondo screening clouds created by two Kondo impurities.
We demonstrate this methodology in the presence of noisy data, showing
the potential of our approach in an experimentally realistic scenario.
This paper is organized as follows. In Sec.~\ref{sec:model} we introduce the Kondo impurity model and formulation of correlation entropy density. 
In Sec.~\ref{sec:ML_model}, we analyze the developed machine learning methodology
to predict the correlation entropy density from measurable local correlators, including its transferability and
the impact of noise.
Finally, in Sec.~\ref{sec:conclusion} we summarize our conclusions. 

\section{Model\label{sec:model}}

We consider the setup shown in Fig.~\ref{fig:setup}(a). Two interacting Kondo spins are are coupled to the opposite sides of a non-interacting
electron gas through Kondo coupling. These couplings induce many-body correlations along the non-interacting gas, 
at a length scale determined by the Kondo cloud length, and when the Kondo clouds overlap they lead to entanglement
between distant Kondo sites. The Hamiltonian of the setup is written as follows
\begin{equation}\label{eq:Hamiltonian}
\begin{split}
&H=\left[J_{k0}\boldsymbol{S}_0\cdot\boldsymbol{s}_1+U\left(\hat{n}_{0\uparrow}-\frac{1}{2}\right)\left(\hat{n}_{0\downarrow}-\frac{1}{2}\right)\right]\\
&-t\sum_{j=1,\sigma}^{n-1}\left(c_{j\sigma}^{\dagger}c_{j+1\sigma}+{\rm{h.c.}}\right)+\mu\sum_{j=1,\sigma}^{n-1}c_{j\sigma}^{\dagger}c_{j\sigma}\\
&+t^{\prime}\sum_{j=1,\sigma}^{n-1}\left(c_{j\sigma}^{\dagger}c_{j+2\sigma}+{\rm{h.c.}}\right)+\bigg[J_{kn}\boldsymbol{s}_n\cdot\boldsymbol{S}_{n+1} \\
&\left.+U\left(\hat{n}_{n+1,\uparrow}-\frac{1}{2}\right)\left(\hat{n}_{n+1,\downarrow}-\frac{1}{2}\right)\right],
\end{split}
\end{equation}
where $c_{j\sigma}^{(\dagger)}$ is the annihilation (creation) operator at site $j$ with spin $\sigma$, $\hat{n}_{j\sigma}$ is the density operator at site $j$ with spin $\sigma$. For the interacting terms $J_{k0}$ and $J_{kn}$ are the Kondo coupling strengths,
$U$ is the on-site interaction to induce charge localization in the Kondo site, 
$\boldsymbol{S}_{0,n+1}$ is the spin-1/2 operator, and $\boldsymbol{s}_{1,n}$ is the local spin operator of the non-interacting chain. For the non-interacting chain, we consider nearest and next nearest-neighbour hopping $t$ and $t'$, and the chemical potential $\mu$. 
We will focus on values of the Kondo couplings corresponding to Kondo clouds smaller than the size of the non-interacting electron gas to minimize finite size effects. 

\begin{figure}[t]
\includegraphics[width=1.0\linewidth]{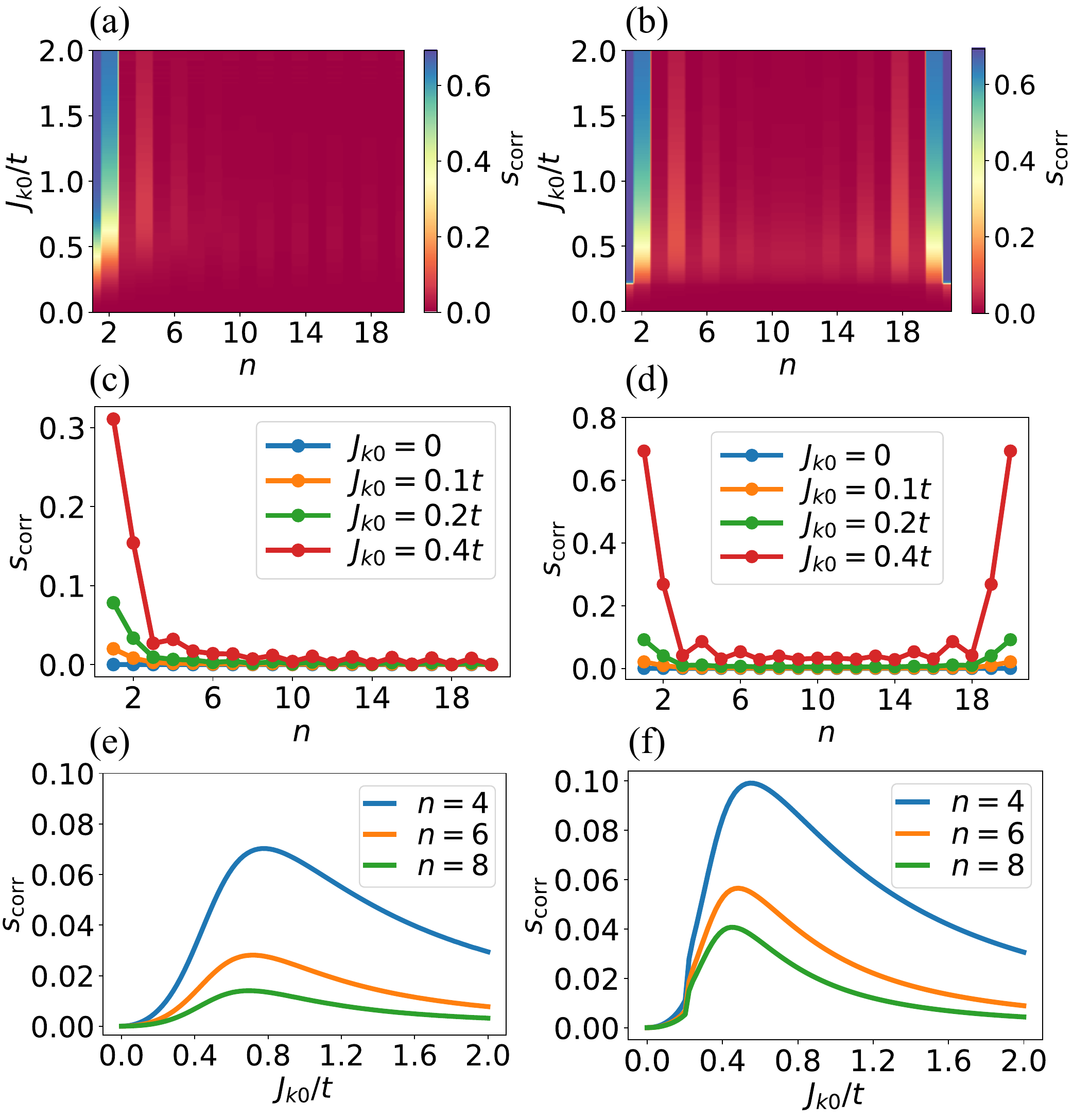}%
\caption{\label{fig:physics_part} Correlation entropy density $s_{\rm{corr}}$ of a 20-site (non-interacting 18-site) model. (a) A heat plot for the case of the interacting spin is located on the left of the non-interacting chain. (b) A heat plot for the case of two interacting spins (with coupling constant $J_{k0}=J_{kn}$) are located on the two sides of the non-interacting chain. (c) Examples of (a) for specific values of the coupling constant $J_{k0}$. (d) Examples of (b) for specific values of the coupling constant $J_{k0}$. (e) Dependence of $s_{\rm{corr}}$ on $J_{k0}$ at different sites in the case of single interacting spin. (f) Dependence of $s_{\rm{corr}}$ on $J_{k0}=J_{kn}$ at different sites in the case of two interacting spin. }
\end{figure}

To characterize the entanglement, we employ the one-particle density matrix, also known as the correlation matrix~\cite{He2014,Lu2014,Fishman2015}. It provides information of the distribution of electrons and their correlations in the system~\cite{Wang2007,Tichy2011,Hofer2013,Esquivel2015}, and is defined as
\begin{equation}\label{eq:ppcorrelator}
C_{ij}^{ss'}
=
\left\langle\Psi_0\right.\vert c_{is}^{\dagger}c_{js'}\left.\vert\Psi_0\right\rangle, 
\end{equation}
where $|\Psi_0\rangle$ refers to a fermionic many-body state. Its eigensolutions offer crucial information about the correlation effect in the many-body state. The eigenvectors $v_k$ defines a set of natural orbitals~\cite{Lwdin1955,Siegbahn1981}, and the corresponding eigenvalues $0\leq\alpha_k\leq 1$ are their ground-state occupation numbers. 
The existence of natural orbitals with eigenvalues  $0<\alpha_k <1$ signifies electronic entanglement. 
Filled and empty orbitals are associated with occupation numbers of 1 and 0, and these orbitals do not contribute to the mode entanglement. 
In the Kondo impurity models, despite the near-macroscopic reorganization of the Fermi sea, the entanglement in the Kondo problem has a few-body character with only a handful of natural orbitals with eigenvalues that significantly differ from 0 and 1~\cite{Yang2017,Zheng2020,Debertolis2021}.
Considering the spatial feature of the induced correlation, we define the correlation entropy density as follows
\begin{equation}\label{eq:definition_scorr}
    s_{\rm{corr}} (\mathbf r) =-\sum_{k}\left(\alpha_k\log\alpha_k \right)\vert v_k\vert ^2 (\mathbf r),
\end{equation}
where $\alpha_k$ are the eigenvalues and the $v_k$ are the corresponding eigenvectors of the one-body density matrix. The full correlation entropy can be determined by integrating over the correlation entropy density $s_{\rm{corr}}$ through the entire fermionic chain. The correlation entropy density serves as a valuable tool for understanding the interaction-induced many-particle correlations within the system. In the absence of particle-particle interactions, all the natural orbitals are either completely filled or empty, and the correlation entropy density $s_{\rm{corr}}$ vanishes. The orbitals with fractional population give rise to a finite $s_{\rm{corr}}$, which also encode the spatial structure of the correlations through $v_k$. 

The correlation entropy density $s_{\rm{corr}}$ of a 20-site fermionic chain, with 18-site non-interacting sites, is shown in Fig.~\ref{fig:physics_part}. The ground state of such a chain is determined by using the tensor-network matrix-product state formalism~\cite{PhysRevLett.69.2863,ITensorurl,itensor,DMRGpyLibrary,zenodolink}, which allows extracting the different particle-particle correlators in the full system and evaluate the correlation entropy density. The correlation entropy density with one interacting spin for various strengths of the coupling constant $J_{k0}$ is shown in Fig.~\ref{fig:physics_part}(a). We can see that $s_{\rm{corr}}$ is strongest at the interacting spin site, and gradually reduces towards to center of the non-interacting chain. The oscillation of $s_{\rm{corr}}$ originates from the oscillation of the particle-particle correlators within a scale of the order of the Fermi wavelength. The horizontal and vertical cuts of Fig.~\ref{fig:physics_part}(a) for specific values of $J_{k0}$ and for specific sites are shown in Fig.~\ref{fig:physics_part}(c) and (e) separately. 

For the case of two interacting spin, the correlation entropy density $s_{\rm{corr}}$ for various strengths of the coupling constant are shown in Fig.~\ref{fig:physics_part}(b,d,f). As one can see, the correlation induced by the other interacting spin brings changes to the profile of $s_{\rm{corr}}$. Two sources of correlation in the non-interacting chain enhances $s_{\rm{corr}}$ throughout the chain. The decay of $s_{\rm{corr}}$ towards the center of the non-interacting chain is also slower than the case of single interacting spin. As the correlation entropy represents the complexity of the correlation, the case of two interacting spins could provide more insights for the quantum entanglement in such systems. 

\section{Machine learning correlation entropy density\label{sec:ML_model}}

\subsection{Local prediction of the correlation density}

We first note that the straightforward experimental extraction of the correlation entropy density $s_{\rm{corr}}$ requires measurement of all the particle-particle correlators. For an $n$-site system, the number of associated correlators is $2n^2$ including long-range ones, 
a greatly challenging task for large systems. This limitation can be bypassed
by directly using a machine learning model to extract the correlation entropy from a reduced set of local correlators. 
In particular, we extract particle-particle correlators related to each specific site by providing local correlators
of the three sites for around each location~\cite{appendix}. At first glance, such an approach leads to a significant information loss, as 
all the non-local correlations required to extract the correaltion entropy
are lost. This information loss is compensated by providing the local
density-density correlators
\begin{equation}\label{eq:ddcorrelator}
    f_{ij}^{ss'}=\left\langle\Psi_0\right.\vert n_{is}n_{js'} \left.\vert\Psi_0\right\rangle
\end{equation}
of the three neighboring sites. The inclusion of density-density correlators provides further information that the conventional
calculation of the correlation entropy does not have access to, but that our machine learning algorithm can exploit to reconstructuct the
correlation entropy.
We will show that this local particle-particle and density-density
correlators are enough to train a supervised learning algorithm to predict the related correlation entropy density. 

\begin{figure}[t]
\includegraphics[width=1.0\linewidth]{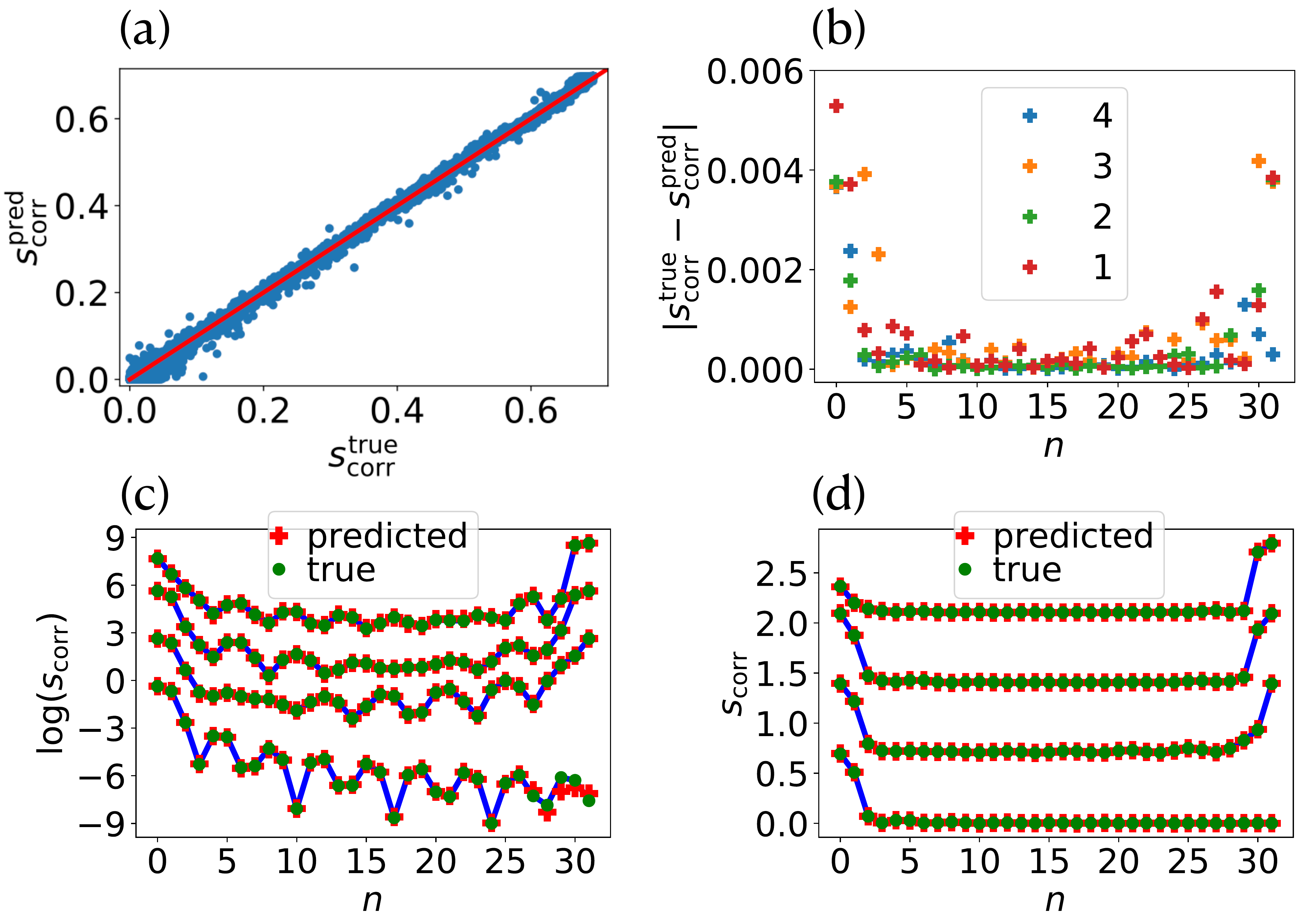}%
\caption{\label{fig:model_evaluation} (a) Comparison between prediction and true values of $s_{\rm{corr}}$. (b) Mean absolute error of $s_{\rm{corr}}$ on each site of four random fermionic chains. (c) Prediction on each site of the entire 32-site chain of $s_{\rm{corr}}$ in log scale for the four fermionic chains in (b). (d) Prediction on each site of the entire 32-site chain of $s_{\rm{corr}}$ for the four fermionic chains in (b).
Curves in (c) (d) are shifted along the vertical axis for clarity. 
}
\end{figure}

As an input, our algorithm assumes correlators around one site and outputs the entropy density at that site.
The training data for the machine learning model is generated according to the following prescription. Solving the Hamiltonian in Eq.~\eqref{eq:Hamiltonian} with randomly generated tight-binding parameters ($t'$, $\mu$, $J_{k0}$, and $J_{kn}$) for a 32-site model enables us to compute the correlation entropy density in Eq.~\eqref{eq:definition_scorr}, which is the quantity to be predicted by the algorithm, exactly at each site. 
The input of the machine learning algorithm could be obtained by measuring the relevant particle-particle correlators in Eq.~\eqref{eq:ppcorrelator} and density-density correlators in Eq.~\eqref{eq:ddcorrelator}. The input correlators of the algorithms correspond to the average between the nth preceding and subsequent sites around a specific site of the fermionic chain~\footnote{For edge sites the average is replaced for the single existing nth neighbor}. This leads to a 
32-dimensional entry for predicting the correlation entropy density. We collected 40,000 examples for training purposes~\cite{appendix}. {The examples are generated with the following ranges of the tight-binding parameters $J_{k0},J_{kn}\in[0,2t]$, $\mu\in[0,2t]$, $t^{\prime}\in[0,t]$. }
We use a Principal Component Analysis~\cite{Jolliffe2016}, keeping all components, to transform the data to an uncorrelated basis. 
For training purposes we use the Box-Cox transformation~\cite{Daimon2011} to
reduce the potential large relative errors created from the small values of $s_{\rm{corr}}$. 
With the transformed dataset, we develop the neural-network structure containing 12 hidden layers with 512 nodes as shown in Fig.~\ref{fig:setup}(b). 

 
The comparison between the predicted and the true values of $s_{\rm{corr}}$ is shown in Fig.~\ref{fig:model_evaluation}(a). The mean absolute error (MAE) of the model is $0.001$. The trained algorithm allows us to predict the correlation entropy density at any site of a model for any sets of tight-binding parameters. In  Fig.~\ref{fig:model_evaluation}(b), the MAE is shown for each site of the fermionic chain for four random fermionic chains. The error values remain at the same levels for the majority sites. Prediction of the each site of four random fermionic chains are shown in Fig.~\ref{fig:model_evaluation}(c) in log scale and Fig.~\ref{fig:model_evaluation}(d) in the original scale. As it can be seen, the predictions match very well with the values of $s_{\rm{corr}}$. 

\begin{figure}[t]
\includegraphics[width=1.0\linewidth]{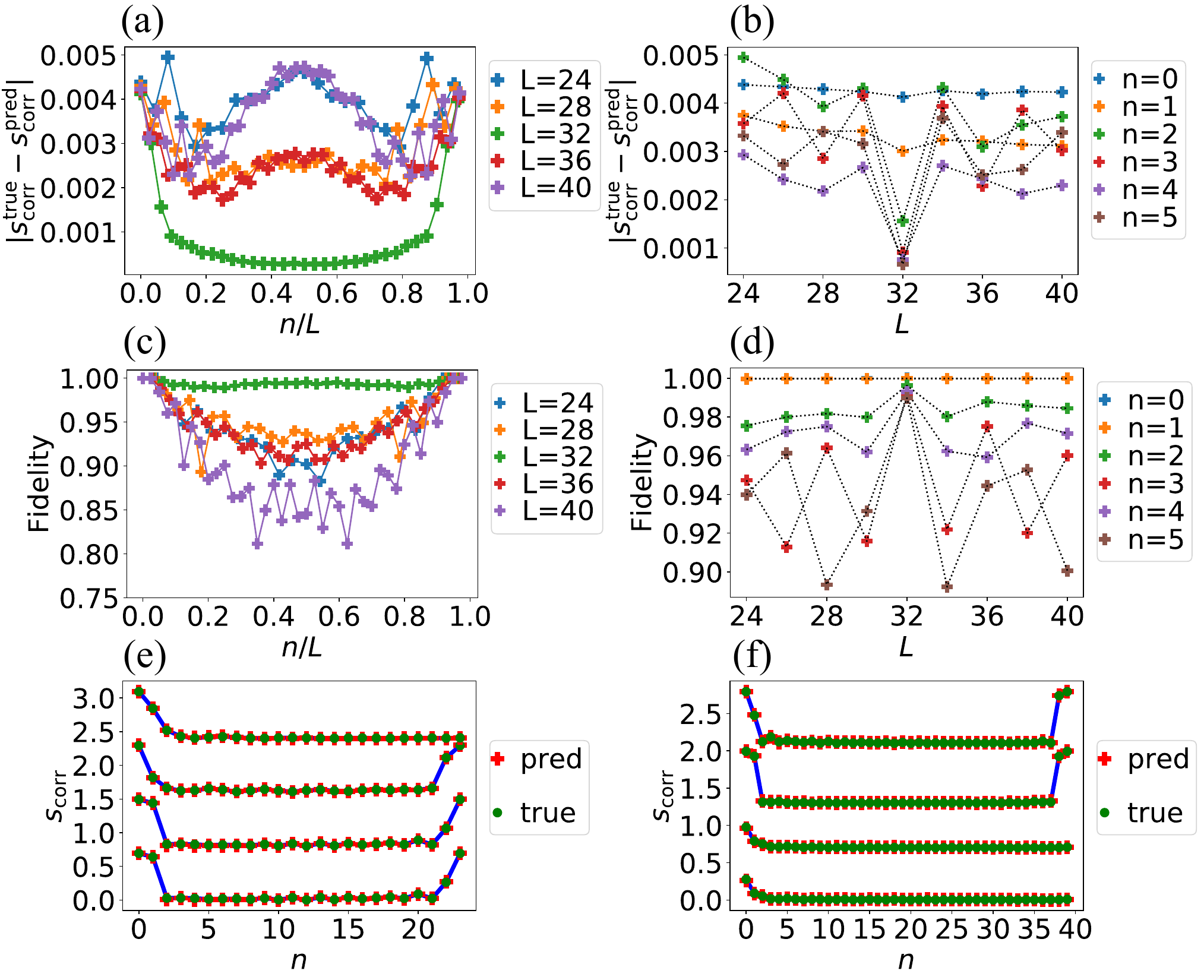}%
\caption{\label{fig:size_independence} Site specific absolute error (a) and fidelity (c) of the prediction of the neural-network model on fermionic chains with different sizes. Size specific absolute error (b) and fidelity (d) of the prediction of the neural-network model on different sites of the fermionic chains. Comparison with the true and prediction on 24-site fermionic chain (e) and on 40-site fermionic chain (f). For clarity the curves are shifted along the vertical axis in (e) and (f).}
\end{figure}

\subsection{Transfer learning to various-size Kondo models\label{sec:site_independence}}
In the following we show how an algorithm training on specific system size allows to make prediction for Kondo models of other systems sizes.
The trained neural-network model, uses local correlators to predict the correlation entropy density $s_{\rm{corr}}$. For a random site, the relevant correlators are only associated with the preceding and subsequent three sites, meaning that the machine learning methodology is local by definition. The model was trained before on a 32-site fermionic chain, but for larger and smaller chains 
the relevant correlators are expected to show an analogous phenomenology for larger and smaller chains. 
This built-in locality in the machine learning algorithm motivates analyzing the potential transferability
of the neural-network model. For this purpose, we directly evaluate the trained neural-network model to predict $s_{\rm{corr}}$ on larger or smaller fermionic chain. We apply the trained neural-network model on 24-, 28-, 36-, and 40- sites fermionic chains, each consisting of 5000 randomly generated examples. 

The site specific MAE for $s_{\rm{corr}}$ of the fermionic chains with different sizes are shown in Fig.~\ref{fig:size_independence}(a). The size specific MAE for $s_{\rm{corr}}$ of the fermionic chains at different sites are shown in Fig.~\ref{fig:size_independence}(b). As can be seen, the average MAE gradually increases for the larger and smaller fermionic chains, but it remains in the error range of the 32-site fermionic chain. Hence, the prediction is reliable for fermionic chains with different sizes. 

The accuracy of the model can also be examined by fidelity defined as
\begin{equation}
\mathcal{F} = 
    \frac{\big\vert \left\langle s_{\rm{corr}}^{\rm{pred}}\cdot s_{\rm{corr}}^{\rm{true}}\right\rangle -\left\langle s_{\rm{corr}}^{\rm{pred}}\right\rangle \cdot \left\langle s_{\rm{corr}}^{\rm{true}}\right\rangle \big\vert}{\sqrt{ \left[\big\langle \left(s_{\rm{corr}}^{\rm{true}}\right)^2\big\rangle-\left\langle s_{\rm{corr}}^{\rm{true}}\right\rangle^2\right] \left[\big\langle \big(s_{\rm{corr}}^{\rm{pred}}\big)^2\big\rangle-\big\langle s_{\rm{corr}}^{\rm{pred}}\big\rangle^2\right] }}
\end{equation}
The previous quantity factors out the impact of different magnitudes fo the correlation entropy
when computing the error, leading to $\mathcal{F}=1$ is the prediction of the 
machine learning algorithm is flawless and $\mathcal{F}=0$ if the algorithm does not have predictive power.
The fidelity for different system sizes and sites is plotted in Fig.~\ref{fig:size_independence}(c) and (d) respectively. 
We observe that the fidelity of the larger and smaller chains also remains in the same range,
despite the algorithm not having been trained in those systems, and the neural-network model is reliable on predicting different sizes of chains. 
As specific examples, the prediction of the each site of four random examples of 24- and 40-sites chains are shown in Fig.~\ref{fig:size_independence}(e) and (f) separately. 
The small departures in the transfer learning can be associated to slightly different finite size effects in the different systems.
Overall, these results demonstrate that a machine learning algorithm of the correlation entropy based on local correlators
is transferable between different systems sizes, further supporting the fact that the correlation entropy can be determined locally.

\subsection{Resilience to noise\label{sec:noise}}

In real experimental data, the extracted correlators may contain a certain amount of noise. For this purpose, we now
address the robustness of the neural-network model by including random numerical noise in the data. 

\begin{figure}[t]
\includegraphics[width=1.0\linewidth]{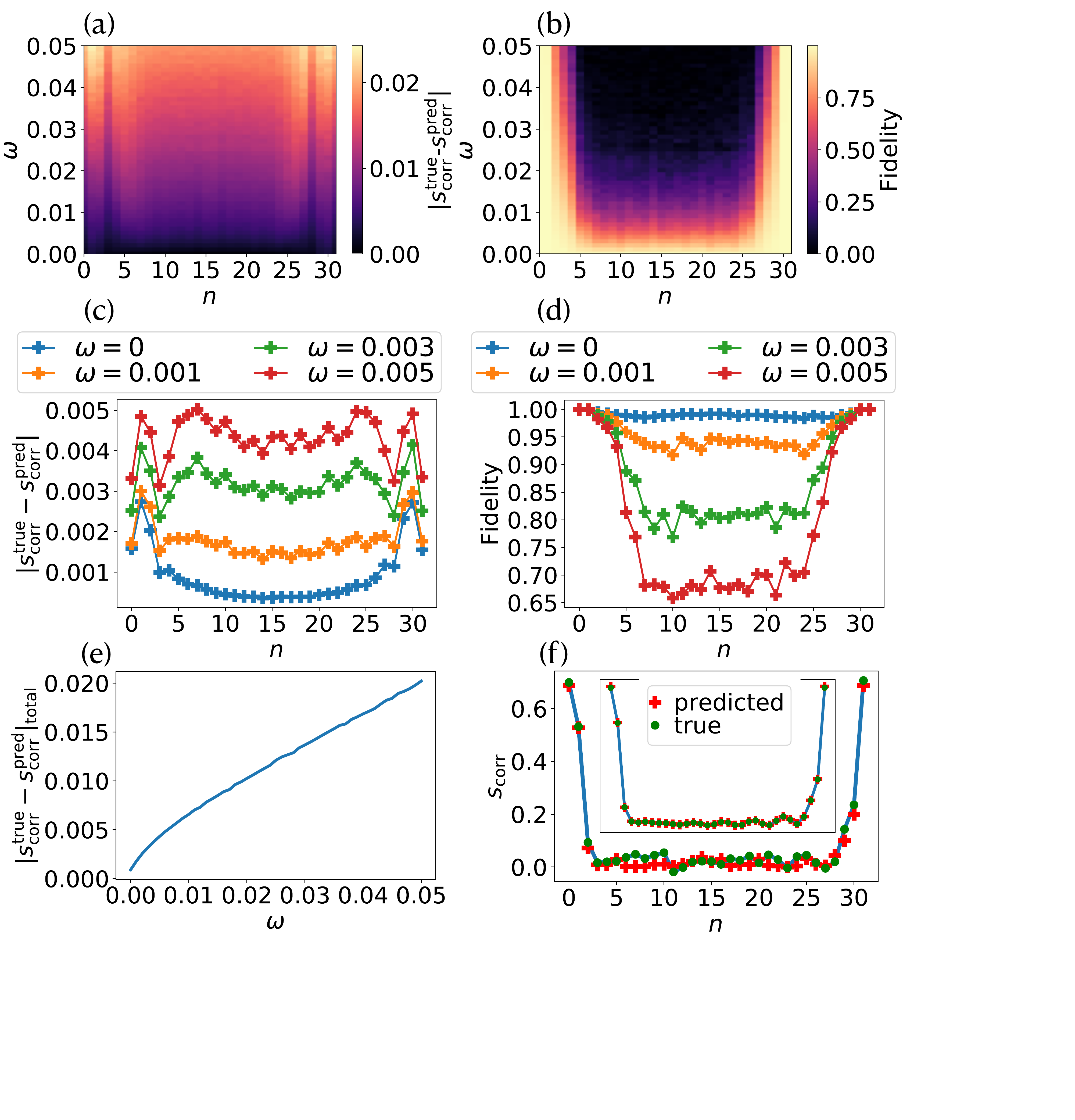}%
\caption{\label{fig:noise} Effect of numerical noise on the prediction of the correlation entropy density trained on the 32-site fermionic chain. (a) Site-specific MAE of the correlation entropy density for various values of the noise rate $\omega$. (b) Site-specific fidelity of the correlation entropy density for various values of the noise rate $\omega$. (c) Horizontal cuts from (a) for smaller values of $\omega$. (d) Horizontal cuts from (b) for smaller values of $\omega$. (e) Total MAE as a function of the noise rate $\omega$.  (f) A comparison between true and predicted values of $s_{\rm{corr}}$ for larger values of $\omega=0.05$ and $\omega=0$ (inset). 
}
\end{figure}

We denote the particle-particle and density-density correlators as $\Lambda_{ij}^{ss',0}=\{C_{ij}^{ss'},f_{ij}^{ss'}\}$, and
introduce the noise in the correlators as 
\begin{equation}
    \Lambda_{ij}^{ss'}=\Lambda_{ij}^{ss',0}+\chi_{ij}^{ss'},
\end{equation}
where $\chi_{ij}^{ss'}$ is the random noise between $[-\omega,\omega]$, and $\omega$ controls the amplitude of the noise. The neural-network model is trained on the 32-site fermionic chain for various degrees of noise and tested for 5000 randomly generated fermionic chains. The MAE and the fidelity are shown in Fig.~\ref{fig:noise}(a) and Fig.~\ref{fig:noise}(b), respectively. 
It is observed that while the prediction of the correlation entropy close to the Kondo impurity is relatively robust,
prediction of the correlation entropy density far from the impurity requires accurate measurements of the correlators.
The error in the correlation density for specific values of $\omega<0.01$ is shown in Fig.~\ref{fig:noise}(c),
and its associated fidelity in Fig.~\ref{fig:noise}(d).
Analogous to Fig.~\ref{fig:noise}(a,b), it is observed that the entropy around the Kondo impurity can be
predicted accurately, whereas far from it the existence of noise decreases the fidelity of the prediction.
The total MAE as a function of $\omega$ as shown in Fig.~\ref{fig:noise}(e), where it is observed that the error increases approximately linearly
with the noise level. As a reference, we show a comparison between the true and predicted $s_{\rm{corr}}$ for $\omega=0.05$ and $\omega=0$ in Fig.~\ref{fig:noise}(f). 
Our results suggest that predicting the correlation entropy featuring low levels of correlation require precise correlator data.
In contrast, predictions of the entropy close to the impurities, which in our calculation corresponds
to a length comparable to the Kondo cloud, are robust to the presence of noise.

\section{Conclusion\label{sec:conclusion}}
In this work, we demonstrated that a machine learning algorithm, assuming local correlators as an input, can accurately predict the many-body entanglement structure of a Kondo screening cloud as characterized by the correlation entropy.
Our methodology combines local single-particle and density correlators, showing that these quantities contain enough information
to reconstruct the correlation entropy in real space. Our method demonstrates that machine learning allows bypassing 
the need to obtain long-range correlators, required for direct methods. 
We showed that, owing to the local nature of the input data, our algorithm is transferable to different system sizes. Thus, our methodology can be applied to systems not included in the training set. We finally demonstrated the resilience of our algorithm to noise, showing that the correlation
entropy is reasonably robust in the presence of sizable inaccuracies in the measured correlators.
The extraction of real-space entanglement offers valuable new insight into the intricate interplay of correlations within the system, including the determination of a spatial profile of the Kondo cloud. 
Our results establish the potential of machine learning methods
to reveal entanglement in many-body systems, including spatially inhomogeneous quantum materials.

\textbf{Acknowledgements} 
FA acknowledges the financial support from the Magnus Ehrnrooth Foundation. 
T.O. acknowledges the Academy of Finland project 331094 for support. 
JLL acknowledges the 
financial support from the
Academy of Finland Projects No. 331342 and No. 358088
and the Jane and Aatos Erkko Foundation.
FA and JLL acknowledge the computational resources provided by
the Aalto Science-IT project.

\appendix

\section{Choice of local correlators and sample size}
\label{sec:appendix}

Here, we address the accuracy of our algorithm as a function of the number of neighboring sites from which correlators were extracted, and the sample size of the training set.

\begin{figure}[t]
\begin{center}
\includegraphics[width=1\linewidth]{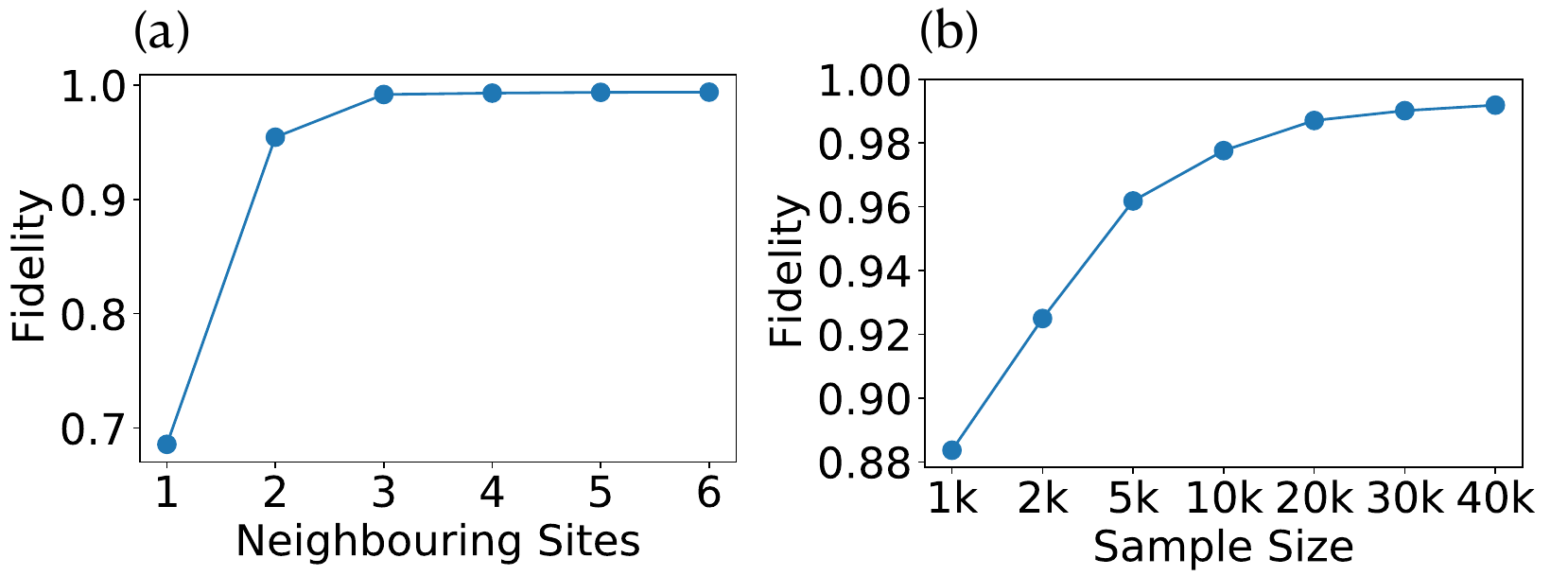}%
\caption{\label{fig:appndix} (a) Fidelity of the neural-network model 
as a function of nearest neighbor sites considered in the training. The maximum sample size 40000 is used in each case.
It is observed that considering correlators in three neighboring sites provides high quality predictions. (b) Fidelity of the neural-network model as a function of the sample size of the training data, the correlators are extracted from three neighbouring sites. It is observed that the accuracy of the algorithm saturates approximately at 20000 samples.}
\end{center}
\end{figure}

Figure \ref{fig:appndix}(a) shows the fidelity $\mathcal{F}$ of the neural-network model trained on correlators extracted from various numbers of neighbouring sites surrounding a given internal site of the non-interacting chain. As it is observed, the accuracy of the model is much lower for neighbouring sites less than three, and improves as the number of neighboring sites is increased. Increasing the number of correlators included significantly increases the number of correlators that must be determined. 
We find that that satisfactory accuracy is attained with correlators extracted from three neighbouring sites, and therefore we focus on the three-neighbor case for the training the optimal neural-network model.

Figure \ref{fig:appndix}(b) shows the fidelity of the neural-network model as a function of the size of the training set. As it is observed, the accuracy of the model increases as the size of the training set increases. We find that the accuracy of the model saturates for sample sizes larger than 20000.
Our calculations are therefore in the regime where the training data is large enough to saturate the accuracy of the algorithm.

Given the behavior described above, we focus the results of our manuscript on a training set with 40000 examples, where the correlators are extracted from three neighboring sites. This choice enables have a modest number of correlators to be determined while maintaining satisfactory accuracy of the neural-network model.

\bibliography{apssamp}{}

\begin{thebibliography}{65}%
\makeatletter
\providecommand \@ifxundefined [1]{%
 \@ifx{#1\undefined}
}%
\providecommand \@ifnum [1]{%
 \ifnum #1\expandafter \@firstoftwo
 \else \expandafter \@secondoftwo
 \fi
}%
\providecommand \@ifx [1]{%
 \ifx #1\expandafter \@firstoftwo
 \else \expandafter \@secondoftwo
 \fi
}%
\providecommand \natexlab [1]{#1}%
\providecommand \enquote  [1]{``#1''}%
\providecommand \bibnamefont  [1]{#1}%
\providecommand \bibfnamefont [1]{#1}%
\providecommand \citenamefont [1]{#1}%
\providecommand \href@noop [0]{\@secondoftwo}%
\providecommand \href [0]{\begingroup \@sanitize@url \@href}%
\providecommand \@href[1]{\@@startlink{#1}\@@href}%
\providecommand \@@href[1]{\endgroup#1\@@endlink}%
\providecommand \@sanitize@url [0]{\catcode `\\12\catcode `\$12\catcode
  `\&12\catcode `\#12\catcode `\^12\catcode `\_12\catcode `\%12\relax}%
\providecommand \@@startlink[1]{}%
\providecommand \@@endlink[0]{}%
\providecommand \url  [0]{\begingroup\@sanitize@url \@url }%
\providecommand \@url [1]{\endgroup\@href {#1}{\urlprefix }}%
\providecommand \urlprefix  [0]{URL }%
\providecommand \Eprint [0]{\href }%
\providecommand \doibase [0]{https://doi.org/}%
\providecommand \selectlanguage [0]{\@gobble}%
\providecommand \bibinfo  [0]{\@secondoftwo}%
\providecommand \bibfield  [0]{\@secondoftwo}%
\providecommand \translation [1]{[#1]}%
\providecommand \BibitemOpen [0]{}%
\providecommand \bibitemStop [0]{}%
\providecommand \bibitemNoStop [0]{.\EOS\space}%
\providecommand \EOS [0]{\spacefactor3000\relax}%
\providecommand \BibitemShut  [1]{\csname bibitem#1\endcsname}%
\let\auto@bib@innerbib\@empty
\bibitem [{\citenamefont {de~Haas}\ \emph {et~al.}(1934)\citenamefont
  {de~Haas}, \citenamefont {de~Boer},\ and\ \citenamefont {van~dën
  Berg}}]{deHaas1934}%
  \BibitemOpen
  \bibfield  {author} {\bibinfo {author} {\bibfnamefont {W.}~\bibnamefont
  {de~Haas}}, \bibinfo {author} {\bibfnamefont {J.}~\bibnamefont {de~Boer}},\
  and\ \bibinfo {author} {\bibfnamefont {G.}~\bibnamefont {van~dën Berg}},\
  }\bibfield  {title} {\bibinfo {title} {The electrical resistance of gold,
  copper and lead at low temperatures},\ }\href
  {https://doi.org/10.1016/s0031-8914(34)80310-2} {\bibfield  {journal}
  {\bibinfo  {journal} {Physica}\ }\textbf {\bibinfo {volume} {1}},\ \bibinfo
  {pages} {1115} (\bibinfo {year} {1934})}\BibitemShut {NoStop}%
\bibitem [{\citenamefont {Kondo}(1964)}]{Kondo1964}%
  \BibitemOpen
  \bibfield  {author} {\bibinfo {author} {\bibfnamefont {J.}~\bibnamefont
  {Kondo}},\ }\bibfield  {title} {\bibinfo {title} {Resistance minimum in
  dilute magnetic alloys},\ }\href {https://doi.org/10.1143/ptp.32.37}
  {\bibfield  {journal} {\bibinfo  {journal} {Progress of Theoretical Physics}\
  }\textbf {\bibinfo {volume} {32}},\ \bibinfo {pages} {37} (\bibinfo {year}
  {1964})}\BibitemShut {NoStop}%
\bibitem [{\citenamefont {Andrei}\ \emph {et~al.}(1983)\citenamefont {Andrei},
  \citenamefont {Furuya},\ and\ \citenamefont
  {Lowenstein}}]{RevModPhys.55.331}%
  \BibitemOpen
  \bibfield  {author} {\bibinfo {author} {\bibfnamefont {N.}~\bibnamefont
  {Andrei}}, \bibinfo {author} {\bibfnamefont {K.}~\bibnamefont {Furuya}},\
  and\ \bibinfo {author} {\bibfnamefont {J.~H.}\ \bibnamefont {Lowenstein}},\
  }\bibfield  {title} {\bibinfo {title} {Solution of the kondo problem},\
  }\href {https://doi.org/10.1103/RevModPhys.55.331} {\bibfield  {journal}
  {\bibinfo  {journal} {Rev. Mod. Phys.}\ }\textbf {\bibinfo {volume} {55}},\
  \bibinfo {pages} {331} (\bibinfo {year} {1983})}\BibitemShut {NoStop}%
\bibitem [{\citenamefont {Hewson}(1993)}]{Hewson1993}%
  \BibitemOpen
  \bibfield  {author} {\bibinfo {author} {\bibfnamefont {A.~C.}\ \bibnamefont
  {Hewson}},\ }\href {https://doi.org/10.1017/cbo9780511470752} {\emph
  {\bibinfo {title} {The Kondo Problem to Heavy Fermions}}}\ (\bibinfo
  {publisher} {Cambridge University Press},\ \bibinfo {year}
  {1993})\BibitemShut {NoStop}%
\bibitem [{\citenamefont {\'Ujs\'aghy}\ \emph {et~al.}(2000)\citenamefont
  {\'Ujs\'aghy}, \citenamefont {Kroha}, \citenamefont {Szunyogh},\ and\
  \citenamefont {Zawadowski}}]{PhysRevLett.85.2557}%
  \BibitemOpen
  \bibfield  {author} {\bibinfo {author} {\bibfnamefont {O.}~\bibnamefont
  {\'Ujs\'aghy}}, \bibinfo {author} {\bibfnamefont {J.}~\bibnamefont {Kroha}},
  \bibinfo {author} {\bibfnamefont {L.}~\bibnamefont {Szunyogh}},\ and\
  \bibinfo {author} {\bibfnamefont {A.}~\bibnamefont {Zawadowski}},\ }\bibfield
   {title} {\bibinfo {title} {Theory of the fano resonance in the stm tunneling
  density of states due to a single kondo impurity},\ }\href
  {https://doi.org/10.1103/PhysRevLett.85.2557} {\bibfield  {journal} {\bibinfo
   {journal} {Phys. Rev. Lett.}\ }\textbf {\bibinfo {volume} {85}},\ \bibinfo
  {pages} {2557} (\bibinfo {year} {2000})}\BibitemShut {NoStop}%
\bibitem [{\citenamefont {Knorr}\ \emph {et~al.}(2002)\citenamefont {Knorr},
  \citenamefont {Schneider}, \citenamefont {Diekh\"oner}, \citenamefont
  {Wahl},\ and\ \citenamefont {Kern}}]{PhysRevLett.88.096804}%
  \BibitemOpen
  \bibfield  {author} {\bibinfo {author} {\bibfnamefont {N.}~\bibnamefont
  {Knorr}}, \bibinfo {author} {\bibfnamefont {M.~A.}\ \bibnamefont
  {Schneider}}, \bibinfo {author} {\bibfnamefont {L.}~\bibnamefont
  {Diekh\"oner}}, \bibinfo {author} {\bibfnamefont {P.}~\bibnamefont {Wahl}},\
  and\ \bibinfo {author} {\bibfnamefont {K.}~\bibnamefont {Kern}},\ }\bibfield
  {title} {\bibinfo {title} {Kondo effect of single co adatoms on cu
  surfaces},\ }\href {https://doi.org/10.1103/PhysRevLett.88.096804} {\bibfield
   {journal} {\bibinfo  {journal} {Phys. Rev. Lett.}\ }\textbf {\bibinfo
  {volume} {88}},\ \bibinfo {pages} {096804} (\bibinfo {year}
  {2002})}\BibitemShut {NoStop}%
\bibitem [{\citenamefont {Cho}\ and\ \citenamefont {McKenzie}(2006)}]{Cho2006}%
  \BibitemOpen
  \bibfield  {author} {\bibinfo {author} {\bibfnamefont {S.~Y.}\ \bibnamefont
  {Cho}}\ and\ \bibinfo {author} {\bibfnamefont {R.~H.}\ \bibnamefont
  {McKenzie}},\ }\bibfield  {title} {\bibinfo {title} {Quantum entanglement in
  the two-impurity kondo model},\ }\bibfield  {journal} {\bibinfo  {journal}
  {Physical Review A}\ }\textbf {\bibinfo {volume} {73}},\ \href
  {https://doi.org/10.1103/physreva.73.012109} {10.1103/physreva.73.012109}
  (\bibinfo {year} {2006})\BibitemShut {NoStop}%
\bibitem [{\citenamefont {Affleck}\ \emph {et~al.}(2009)\citenamefont
  {Affleck}, \citenamefont {Laflorencie},\ and\ \citenamefont
  {S{\o}rensen}}]{Affleck2009}%
  \BibitemOpen
  \bibfield  {author} {\bibinfo {author} {\bibfnamefont {I.}~\bibnamefont
  {Affleck}}, \bibinfo {author} {\bibfnamefont {N.}~\bibnamefont
  {Laflorencie}},\ and\ \bibinfo {author} {\bibfnamefont {E.~S.}\ \bibnamefont
  {S{\o}rensen}},\ }\bibfield  {title} {\bibinfo {title} {Entanglement entropy
  in quantum impurity systems and systems with boundaries},\ }\href
  {https://doi.org/10.1088/1751-8113/42/50/504009} {\bibfield  {journal}
  {\bibinfo  {journal} {Journal of Physics A: Mathematical and Theoretical}\
  }\textbf {\bibinfo {volume} {42}},\ \bibinfo {pages} {504009} (\bibinfo
  {year} {2009})}\BibitemShut {NoStop}%
\bibitem [{\citenamefont {Wirth}\ and\ \citenamefont
  {Steglich}(2016)}]{Wirth2016}%
  \BibitemOpen
  \bibfield  {author} {\bibinfo {author} {\bibfnamefont {S.}~\bibnamefont
  {Wirth}}\ and\ \bibinfo {author} {\bibfnamefont {F.}~\bibnamefont
  {Steglich}},\ }\bibfield  {title} {\bibinfo {title} {Exploring heavy fermions
  from macroscopic to microscopic length scales},\ }\bibfield  {journal}
  {\bibinfo  {journal} {Nature Reviews Materials}\ }\textbf {\bibinfo {volume}
  {1}},\ \href {https://doi.org/10.1038/natrevmats.2016.51}
  {10.1038/natrevmats.2016.51} (\bibinfo {year} {2016})\BibitemShut {NoStop}%
\bibitem [{\citenamefont {Tsunetsugu}\ \emph {et~al.}(1997)\citenamefont
  {Tsunetsugu}, \citenamefont {Sigrist},\ and\ \citenamefont
  {Ueda}}]{RevModPhys.69.809}%
  \BibitemOpen
  \bibfield  {author} {\bibinfo {author} {\bibfnamefont {H.}~\bibnamefont
  {Tsunetsugu}}, \bibinfo {author} {\bibfnamefont {M.}~\bibnamefont
  {Sigrist}},\ and\ \bibinfo {author} {\bibfnamefont {K.}~\bibnamefont
  {Ueda}},\ }\bibfield  {title} {\bibinfo {title} {The ground-state phase
  diagram of the one-dimensional kondo lattice model},\ }\href
  {https://doi.org/10.1103/RevModPhys.69.809} {\bibfield  {journal} {\bibinfo
  {journal} {Rev. Mod. Phys.}\ }\textbf {\bibinfo {volume} {69}},\ \bibinfo
  {pages} {809} (\bibinfo {year} {1997})}\BibitemShut {NoStop}%
\bibitem [{\citenamefont {Mukherjee}\ \emph {et~al.}(2022)\citenamefont
  {Mukherjee}, \citenamefont {Mukherjee}, \citenamefont {Vidhyadhiraja},
  \citenamefont {Taraphder},\ and\ \citenamefont {Lal}}]{Mukherjee2022}%
  \BibitemOpen
  \bibfield  {author} {\bibinfo {author} {\bibfnamefont {A.}~\bibnamefont
  {Mukherjee}}, \bibinfo {author} {\bibfnamefont {A.}~\bibnamefont
  {Mukherjee}}, \bibinfo {author} {\bibfnamefont {N.~S.}\ \bibnamefont
  {Vidhyadhiraja}}, \bibinfo {author} {\bibfnamefont {A.}~\bibnamefont
  {Taraphder}},\ and\ \bibinfo {author} {\bibfnamefont {S.}~\bibnamefont
  {Lal}},\ }\bibfield  {title} {\bibinfo {title} {Unveiling the kondo cloud:
  Unitary renormalization-group study of the kondo model},\ }\bibfield
  {journal} {\bibinfo  {journal} {Physical Review B}\ }\textbf {\bibinfo
  {volume} {105}},\ \href {https://doi.org/10.1103/physrevb.105.085119}
  {10.1103/physrevb.105.085119} (\bibinfo {year} {2022})\BibitemShut {NoStop}%
\bibitem [{\citenamefont {Barzykin}\ and\ \citenamefont
  {Affleck}(1996)}]{Barzykin1996}%
  \BibitemOpen
  \bibfield  {author} {\bibinfo {author} {\bibfnamefont {V.}~\bibnamefont
  {Barzykin}}\ and\ \bibinfo {author} {\bibfnamefont {I.}~\bibnamefont
  {Affleck}},\ }\bibfield  {title} {\bibinfo {title} {The kondo screening
  cloud: What can we learn from perturbation theory?},\ }\href
  {https://doi.org/10.1103/physrevlett.76.4959} {\bibfield  {journal} {\bibinfo
   {journal} {Physical Review Letters}\ }\textbf {\bibinfo {volume} {76}},\
  \bibinfo {pages} {4959} (\bibinfo {year} {1996})}\BibitemShut {NoStop}%
\bibitem [{\citenamefont {Affleck}(2010)}]{Affleck2010}%
  \BibitemOpen
  \bibfield  {author} {\bibinfo {author} {\bibfnamefont {I.}~\bibnamefont
  {Affleck}},\ }\bibfield  {title} {\bibinfo {title} {The kondo screening
  cloud: What it is and how to observe it},\ }in\ \href
  {https://doi.org/10.1142/9789814299442_0001} {\emph {\bibinfo {booktitle}
  {Perspectives of Mesoscopic Physics}}}\ (\bibinfo  {publisher} {{WORLD}
  {SCIENTIFIC}},\ \bibinfo {year} {2010})\ pp.\ \bibinfo {pages}
  {1--44}\BibitemShut {NoStop}%
\bibitem [{\citenamefont {Borzenets}\ \emph {et~al.}(2020)\citenamefont
  {Borzenets}, \citenamefont {Shim}, \citenamefont {Chen}, \citenamefont
  {Ludwig}, \citenamefont {Wieck}, \citenamefont {Tarucha}, \citenamefont
  {Sim},\ and\ \citenamefont {Yamamoto}}]{VBorzenets2020}%
  \BibitemOpen
  \bibfield  {author} {\bibinfo {author} {\bibfnamefont {I.~V.}\ \bibnamefont
  {Borzenets}}, \bibinfo {author} {\bibfnamefont {J.}~\bibnamefont {Shim}},
  \bibinfo {author} {\bibfnamefont {J.~C.~H.}\ \bibnamefont {Chen}}, \bibinfo
  {author} {\bibfnamefont {A.}~\bibnamefont {Ludwig}}, \bibinfo {author}
  {\bibfnamefont {A.~D.}\ \bibnamefont {Wieck}}, \bibinfo {author}
  {\bibfnamefont {S.}~\bibnamefont {Tarucha}}, \bibinfo {author} {\bibfnamefont
  {H.-S.}\ \bibnamefont {Sim}},\ and\ \bibinfo {author} {\bibfnamefont
  {M.}~\bibnamefont {Yamamoto}},\ }\bibfield  {title} {\bibinfo {title}
  {Observation of the kondo screening cloud},\ }\href
  {https://doi.org/10.1038/s41586-020-2058-6} {\bibfield  {journal} {\bibinfo
  {journal} {Nature}\ }\textbf {\bibinfo {volume} {579}},\ \bibinfo {pages}
  {210} (\bibinfo {year} {2020})}\BibitemShut {NoStop}%
\bibitem [{\citenamefont {Gubernatis}\ \emph {et~al.}(1987)\citenamefont
  {Gubernatis}, \citenamefont {Hirsch},\ and\ \citenamefont
  {Scalapino}}]{Gubernatis1987}%
  \BibitemOpen
  \bibfield  {author} {\bibinfo {author} {\bibfnamefont {J.~E.}\ \bibnamefont
  {Gubernatis}}, \bibinfo {author} {\bibfnamefont {J.~E.}\ \bibnamefont
  {Hirsch}},\ and\ \bibinfo {author} {\bibfnamefont {D.~J.}\ \bibnamefont
  {Scalapino}},\ }\bibfield  {title} {\bibinfo {title} {Spin and charge
  correlations around an anderson magnetic impurity},\ }\href
  {https://doi.org/10.1103/physrevb.35.8478} {\bibfield  {journal} {\bibinfo
  {journal} {Physical Review B}\ }\textbf {\bibinfo {volume} {35}},\ \bibinfo
  {pages} {8478} (\bibinfo {year} {1987})}\BibitemShut {NoStop}%
\bibitem [{\citenamefont {Barzykin}\ and\ \citenamefont
  {Affleck}(1998)}]{Barzykin1998}%
  \BibitemOpen
  \bibfield  {author} {\bibinfo {author} {\bibfnamefont {V.}~\bibnamefont
  {Barzykin}}\ and\ \bibinfo {author} {\bibfnamefont {I.}~\bibnamefont
  {Affleck}},\ }\bibfield  {title} {\bibinfo {title} {Screening cloud in the
  $k$-channel kondo model: Perturbative and large-$k$ results},\ }\href
  {https://doi.org/10.1103/physrevb.57.432} {\bibfield  {journal} {\bibinfo
  {journal} {Physical Review B}\ }\textbf {\bibinfo {volume} {57}},\ \bibinfo
  {pages} {432} (\bibinfo {year} {1998})}\BibitemShut {NoStop}%
\bibitem [{\citenamefont {Borda}(2007)}]{Borda2007}%
  \BibitemOpen
  \bibfield  {author} {\bibinfo {author} {\bibfnamefont {L.}~\bibnamefont
  {Borda}},\ }\bibfield  {title} {\bibinfo {title} {Kondo screening cloud in a
  one-dimensional wire: Numerical renormalization group study},\ }\bibfield
  {journal} {\bibinfo  {journal} {Physical Review B}\ }\textbf {\bibinfo
  {volume} {75}},\ \href {https://doi.org/10.1103/physrevb.75.041307}
  {10.1103/physrevb.75.041307} (\bibinfo {year} {2007})\BibitemShut {NoStop}%
\bibitem [{\citenamefont {Holzner}\ \emph {et~al.}(2009)\citenamefont
  {Holzner}, \citenamefont {McCulloch}, \citenamefont {Schollw\"{o}ck},
  \citenamefont {von Delft},\ and\ \citenamefont
  {Heidrich-Meisner}}]{Holzner2009}%
  \BibitemOpen
  \bibfield  {author} {\bibinfo {author} {\bibfnamefont {A.}~\bibnamefont
  {Holzner}}, \bibinfo {author} {\bibfnamefont {I.~P.}\ \bibnamefont
  {McCulloch}}, \bibinfo {author} {\bibfnamefont {U.}~\bibnamefont
  {Schollw\"{o}ck}}, \bibinfo {author} {\bibfnamefont {J.}~\bibnamefont {von
  Delft}},\ and\ \bibinfo {author} {\bibfnamefont {F.}~\bibnamefont
  {Heidrich-Meisner}},\ }\bibfield  {title} {\bibinfo {title} {Kondo screening
  cloud in the single-impurity anderson model: A density matrix renormalization
  group study},\ }\bibfield  {journal} {\bibinfo  {journal} {Physical Review
  B}\ }\textbf {\bibinfo {volume} {80}},\ \href
  {https://doi.org/10.1103/physrevb.80.205114} {10.1103/physrevb.80.205114}
  (\bibinfo {year} {2009})\BibitemShut {NoStop}%
\bibitem [{\citenamefont {He}\ and\ \citenamefont {Lu}(2014)}]{He2014}%
  \BibitemOpen
  \bibfield  {author} {\bibinfo {author} {\bibfnamefont {R.-Q.}\ \bibnamefont
  {He}}\ and\ \bibinfo {author} {\bibfnamefont {Z.-Y.}\ \bibnamefont {Lu}},\
  }\bibfield  {title} {\bibinfo {title} {Quantum renormalization groups based
  on natural orbitals},\ }\bibfield  {journal} {\bibinfo  {journal} {Physical
  Review B}\ }\textbf {\bibinfo {volume} {89}},\ \href
  {https://doi.org/10.1103/physrevb.89.085108} {10.1103/physrevb.89.085108}
  (\bibinfo {year} {2014})\BibitemShut {NoStop}%
\bibitem [{\citenamefont {Lu}\ \emph {et~al.}(2014)\citenamefont {Lu},
  \citenamefont {H\"{o}ppner}, \citenamefont {Gunnarsson},\ and\ \citenamefont
  {Haverkort}}]{Lu2014}%
  \BibitemOpen
  \bibfield  {author} {\bibinfo {author} {\bibfnamefont {Y.}~\bibnamefont
  {Lu}}, \bibinfo {author} {\bibfnamefont {M.}~\bibnamefont {H\"{o}ppner}},
  \bibinfo {author} {\bibfnamefont {O.}~\bibnamefont {Gunnarsson}},\ and\
  \bibinfo {author} {\bibfnamefont {M.~W.}\ \bibnamefont {Haverkort}},\
  }\bibfield  {title} {\bibinfo {title} {Efficient real-frequency solver for
  dynamical mean-field theory},\ }\bibfield  {journal} {\bibinfo  {journal}
  {Physical Review B}\ }\textbf {\bibinfo {volume} {90}},\ \href
  {https://doi.org/10.1103/physrevb.90.085102} {10.1103/physrevb.90.085102}
  (\bibinfo {year} {2014})\BibitemShut {NoStop}%
\bibitem [{\citenamefont {Fishman}\ and\ \citenamefont
  {White}(2015)}]{Fishman2015}%
  \BibitemOpen
  \bibfield  {author} {\bibinfo {author} {\bibfnamefont {M.~T.}\ \bibnamefont
  {Fishman}}\ and\ \bibinfo {author} {\bibfnamefont {S.~R.}\ \bibnamefont
  {White}},\ }\bibfield  {title} {\bibinfo {title} {Compression of correlation
  matrices and an efficient method for forming matrix product states of
  fermionic gaussian states},\ }\bibfield  {journal} {\bibinfo  {journal}
  {Physical Review B}\ }\textbf {\bibinfo {volume} {92}},\ \href
  {https://doi.org/10.1103/physrevb.92.075132} {10.1103/physrevb.92.075132}
  (\bibinfo {year} {2015})\BibitemShut {NoStop}%
\bibitem [{\citenamefont {Esquivel}\ \emph {et~al.}(1996)\citenamefont
  {Esquivel}, \citenamefont {Rodr{\'{\i}}guez}, \citenamefont {Sagar},
  \citenamefont {H{\^{o}}},\ and\ \citenamefont {Smith}}]{Esquivel1996}%
  \BibitemOpen
  \bibfield  {author} {\bibinfo {author} {\bibfnamefont {R.~O.}\ \bibnamefont
  {Esquivel}}, \bibinfo {author} {\bibfnamefont {A.~L.}\ \bibnamefont
  {Rodr{\'{\i}}guez}}, \bibinfo {author} {\bibfnamefont {R.~P.}\ \bibnamefont
  {Sagar}}, \bibinfo {author} {\bibfnamefont {M.}~\bibnamefont {H{\^{o}}}},\
  and\ \bibinfo {author} {\bibfnamefont {V.~H.}\ \bibnamefont {Smith}},\
  }\bibfield  {title} {\bibinfo {title} {Physical interpretation of information
  entropy: Numerical evidence of the collins conjecture},\ }\href
  {https://doi.org/10.1103/physreva.54.259} {\bibfield  {journal} {\bibinfo
  {journal} {Physical Review A}\ }\textbf {\bibinfo {volume} {54}},\ \bibinfo
  {pages} {259} (\bibinfo {year} {1996})}\BibitemShut {NoStop}%
\bibitem [{\citenamefont {Gersdorf}\ \emph {et~al.}(1997)\citenamefont
  {Gersdorf}, \citenamefont {John}, \citenamefont {Perdew},\ and\ \citenamefont
  {Ziesche}}]{Gersdorf1997}%
  \BibitemOpen
  \bibfield  {author} {\bibinfo {author} {\bibfnamefont {P.}~\bibnamefont
  {Gersdorf}}, \bibinfo {author} {\bibfnamefont {W.}~\bibnamefont {John}},
  \bibinfo {author} {\bibfnamefont {J.~P.}\ \bibnamefont {Perdew}},\ and\
  \bibinfo {author} {\bibfnamefont {P.}~\bibnamefont {Ziesche}},\ }\bibfield
  {title} {\bibinfo {title} {Correlation entropy of the h2 molecule},\ }\href
  {https://doi.org/10.1002/(sici)1097-461x(1997)61:6<935::aid-qua6>3.0.co;2-x}
  {\bibfield  {journal} {\bibinfo  {journal} {International Journal of Quantum
  Chemistry}\ }\textbf {\bibinfo {volume} {61}},\ \bibinfo {pages} {935}
  (\bibinfo {year} {1997})}\BibitemShut {NoStop}%
\bibitem [{\citenamefont {Ziesche}\ \emph {et~al.}(1997)\citenamefont
  {Ziesche}, \citenamefont {Gunnarsson}, \citenamefont {John},\ and\
  \citenamefont {Beck}}]{Ziesche1997}%
  \BibitemOpen
  \bibfield  {author} {\bibinfo {author} {\bibfnamefont {P.}~\bibnamefont
  {Ziesche}}, \bibinfo {author} {\bibfnamefont {O.}~\bibnamefont {Gunnarsson}},
  \bibinfo {author} {\bibfnamefont {W.}~\bibnamefont {John}},\ and\ \bibinfo
  {author} {\bibfnamefont {H.}~\bibnamefont {Beck}},\ }\bibfield  {title}
  {\bibinfo {title} {Two-site hubbard model, the bardeen-cooper-schrieffer
  model, and the concept of correlation entropy},\ }\href
  {https://doi.org/10.1103/physrevb.55.10270} {\bibfield  {journal} {\bibinfo
  {journal} {Physical Review B}\ }\textbf {\bibinfo {volume} {55}},\ \bibinfo
  {pages} {10270} (\bibinfo {year} {1997})}\BibitemShut {NoStop}%
\bibitem [{\citenamefont {Huang}\ \emph {et~al.}(2006)\citenamefont {Huang},
  \citenamefont {Wang},\ and\ \citenamefont {Kais}}]{Huang2006}%
  \BibitemOpen
  \bibfield  {author} {\bibinfo {author} {\bibfnamefont {Z.}~\bibnamefont
  {Huang}}, \bibinfo {author} {\bibfnamefont {H.}~\bibnamefont {Wang}},\ and\
  \bibinfo {author} {\bibfnamefont {S.}~\bibnamefont {Kais}},\ }\bibfield
  {title} {\bibinfo {title} {Entanglement and electron correlation in quantum
  chemistry calculations},\ }\href {https://doi.org/10.1080/09500340600955674}
  {\bibfield  {journal} {\bibinfo  {journal} {Journal of Modern Optics}\
  }\textbf {\bibinfo {volume} {53}},\ \bibinfo {pages} {2543} (\bibinfo {year}
  {2006})}\BibitemShut {NoStop}%
\bibitem [{\citenamefont {Benavides-Riveros}\ \emph {et~al.}(2017)\citenamefont
  {Benavides-Riveros}, \citenamefont {Lathiotakis}, \citenamefont {Schilling},\
  and\ \citenamefont {Marques}}]{BenavidesRiveros2017}%
  \BibitemOpen
  \bibfield  {author} {\bibinfo {author} {\bibfnamefont {C.~L.}\ \bibnamefont
  {Benavides-Riveros}}, \bibinfo {author} {\bibfnamefont {N.~N.}\ \bibnamefont
  {Lathiotakis}}, \bibinfo {author} {\bibfnamefont {C.}~\bibnamefont
  {Schilling}},\ and\ \bibinfo {author} {\bibfnamefont {M.~A.~L.}\ \bibnamefont
  {Marques}},\ }\bibfield  {title} {\bibinfo {title} {Relating correlation
  measures: The importance of the energy gap},\ }\bibfield  {journal} {\bibinfo
   {journal} {Physical Review A}\ }\textbf {\bibinfo {volume} {95}},\ \href
  {https://doi.org/10.1103/physreva.95.032507} {10.1103/physreva.95.032507}
  (\bibinfo {year} {2017})\BibitemShut {NoStop}%
\bibitem [{\citenamefont {Ferreira}\ \emph {et~al.}(2022)\citenamefont
  {Ferreira}, \citenamefont {Maciel}, \citenamefont {Vianna},\ and\
  \citenamefont {Iemini}}]{Ferreira2022}%
  \BibitemOpen
  \bibfield  {author} {\bibinfo {author} {\bibfnamefont {D.~L.~B.}\
  \bibnamefont {Ferreira}}, \bibinfo {author} {\bibfnamefont {T.~O.}\
  \bibnamefont {Maciel}}, \bibinfo {author} {\bibfnamefont {R.~O.}\
  \bibnamefont {Vianna}},\ and\ \bibinfo {author} {\bibfnamefont
  {F.}~\bibnamefont {Iemini}},\ }\bibfield  {title} {\bibinfo {title} {Quantum
  correlations, entanglement spectrum, and coherence of the two-particle
  reduced density matrix in the extended hubbard model},\ }\bibfield  {journal}
  {\bibinfo  {journal} {Physical Review B}\ }\textbf {\bibinfo {volume}
  {105}},\ \href {https://doi.org/10.1103/physrevb.105.115145}
  {10.1103/physrevb.105.115145} (\bibinfo {year} {2022})\BibitemShut {NoStop}%
\bibitem [{\citenamefont {Aikebaier}\ \emph {et~al.}(2023)\citenamefont
  {Aikebaier}, \citenamefont {Ojanen},\ and\ \citenamefont
  {Lado}}]{Aikebaier2023}%
  \BibitemOpen
  \bibfield  {author} {\bibinfo {author} {\bibfnamefont {F.}~\bibnamefont
  {Aikebaier}}, \bibinfo {author} {\bibfnamefont {T.}~\bibnamefont {Ojanen}},\
  and\ \bibinfo {author} {\bibfnamefont {J.~L.}\ \bibnamefont {Lado}},\
  }\bibfield  {title} {\bibinfo {title} {Extracting electronic many-body
  correlations from local measurements with artificial neural networks},\
  }\bibfield  {journal} {\bibinfo  {journal} {{SciPost} Physics Core}\ }\textbf
  {\bibinfo {volume} {6}},\ \href
  {https://doi.org/10.21468/scipostphyscore.6.2.030}
  {10.21468/scipostphyscore.6.2.030} (\bibinfo {year} {2023})\BibitemShut
  {NoStop}%
\bibitem [{\citenamefont {Schirmer}\ and\ \citenamefont
  {Oi}(2009)}]{PhysRevA.80.022333}%
  \BibitemOpen
  \bibfield  {author} {\bibinfo {author} {\bibfnamefont {S.~G.}\ \bibnamefont
  {Schirmer}}\ and\ \bibinfo {author} {\bibfnamefont {D.~K.~L.}\ \bibnamefont
  {Oi}},\ }\bibfield  {title} {\bibinfo {title} {Two-qubit hamiltonian
  tomography by bayesian analysis of noisy data},\ }\href
  {https://doi.org/10.1103/PhysRevA.80.022333} {\bibfield  {journal} {\bibinfo
  {journal} {Phys. Rev. A}\ }\textbf {\bibinfo {volume} {80}},\ \bibinfo
  {pages} {022333} (\bibinfo {year} {2009})}\BibitemShut {NoStop}%
\bibitem [{\citenamefont {Wang}\ \emph {et~al.}(2017)\citenamefont {Wang},
  \citenamefont {Paesani}, \citenamefont {Santagati}, \citenamefont {Knauer},
  \citenamefont {Gentile}, \citenamefont {Wiebe}, \citenamefont {Petruzzella},
  \citenamefont {O'Brien}, \citenamefont {Rarity}, \citenamefont {Laing},\ and\
  \citenamefont {Thompson}}]{Wang2017}%
  \BibitemOpen
  \bibfield  {author} {\bibinfo {author} {\bibfnamefont {J.}~\bibnamefont
  {Wang}}, \bibinfo {author} {\bibfnamefont {S.}~\bibnamefont {Paesani}},
  \bibinfo {author} {\bibfnamefont {R.}~\bibnamefont {Santagati}}, \bibinfo
  {author} {\bibfnamefont {S.}~\bibnamefont {Knauer}}, \bibinfo {author}
  {\bibfnamefont {A.~A.}\ \bibnamefont {Gentile}}, \bibinfo {author}
  {\bibfnamefont {N.}~\bibnamefont {Wiebe}}, \bibinfo {author} {\bibfnamefont
  {M.}~\bibnamefont {Petruzzella}}, \bibinfo {author} {\bibfnamefont {J.~L.}\
  \bibnamefont {O'Brien}}, \bibinfo {author} {\bibfnamefont {J.~G.}\
  \bibnamefont {Rarity}}, \bibinfo {author} {\bibfnamefont {A.}~\bibnamefont
  {Laing}},\ and\ \bibinfo {author} {\bibfnamefont {M.~G.}\ \bibnamefont
  {Thompson}},\ }\bibfield  {title} {\bibinfo {title} {Experimental quantum
  hamiltonian learning},\ }\href {https://doi.org/10.1038/nphys4074} {\bibfield
   {journal} {\bibinfo  {journal} {Nature Physics}\ }\textbf {\bibinfo {volume}
  {13}},\ \bibinfo {pages} {551} (\bibinfo {year} {2017})}\BibitemShut
  {NoStop}%
\bibitem [{\citenamefont {Hincks}\ \emph {et~al.}(2018)\citenamefont {Hincks},
  \citenamefont {Granade},\ and\ \citenamefont {Cory}}]{Hincks2018}%
  \BibitemOpen
  \bibfield  {author} {\bibinfo {author} {\bibfnamefont {I.}~\bibnamefont
  {Hincks}}, \bibinfo {author} {\bibfnamefont {C.}~\bibnamefont {Granade}},\
  and\ \bibinfo {author} {\bibfnamefont {D.~G.}\ \bibnamefont {Cory}},\
  }\bibfield  {title} {\bibinfo {title} {Statistical inference with quantum
  measurements: methodologies for nitrogen vacancy centers in diamond},\ }\href
  {https://doi.org/10.1088/1367-2630/aa9c9f} {\bibfield  {journal} {\bibinfo
  {journal} {New Journal of Physics}\ }\textbf {\bibinfo {volume} {20}},\
  \bibinfo {pages} {013022} (\bibinfo {year} {2018})}\BibitemShut {NoStop}%
\bibitem [{\citenamefont {Karjalainen}\ \emph {et~al.}(2023)\citenamefont
  {Karjalainen}, \citenamefont {Lippo}, \citenamefont {Chen}, \citenamefont
  {Koch}, \citenamefont {Fumega},\ and\ \citenamefont
  {Lado}}]{PhysRevApplied.20.024054}%
  \BibitemOpen
  \bibfield  {author} {\bibinfo {author} {\bibfnamefont {N.}~\bibnamefont
  {Karjalainen}}, \bibinfo {author} {\bibfnamefont {Z.}~\bibnamefont {Lippo}},
  \bibinfo {author} {\bibfnamefont {G.}~\bibnamefont {Chen}}, \bibinfo {author}
  {\bibfnamefont {R.}~\bibnamefont {Koch}}, \bibinfo {author} {\bibfnamefont
  {A.~O.}\ \bibnamefont {Fumega}},\ and\ \bibinfo {author} {\bibfnamefont
  {J.~L.}\ \bibnamefont {Lado}},\ }\bibfield  {title} {\bibinfo {title}
  {Hamiltonian inference from dynamical excitations in confined quantum
  magnets},\ }\href {https://doi.org/10.1103/PhysRevApplied.20.024054}
  {\bibfield  {journal} {\bibinfo  {journal} {Phys. Rev. Appl.}\ }\textbf
  {\bibinfo {volume} {20}},\ \bibinfo {pages} {024054} (\bibinfo {year}
  {2023})}\BibitemShut {NoStop}%
\bibitem [{\citenamefont {Khosravian}\ \emph {et~al.}(2024)\citenamefont
  {Khosravian}, \citenamefont {Koch},\ and\ \citenamefont
  {Lado}}]{Khosravian2024}%
  \BibitemOpen
  \bibfield  {author} {\bibinfo {author} {\bibfnamefont {M.}~\bibnamefont
  {Khosravian}}, \bibinfo {author} {\bibfnamefont {R.}~\bibnamefont {Koch}},\
  and\ \bibinfo {author} {\bibfnamefont {J.~L.}\ \bibnamefont {Lado}},\
  }\bibfield  {title} {\bibinfo {title} {Hamiltonian learning with real-space
  impurity tomography in topological moiré superconductors},\ }\href
  {https://doi.org/10.1088/2515-7639/ad1c04} {\bibfield  {journal} {\bibinfo
  {journal} {Journal of Physics: Materials}\ }\textbf {\bibinfo {volume} {7}},\
  \bibinfo {pages} {015012} (\bibinfo {year} {2024})}\BibitemShut {NoStop}%
\bibitem [{\citenamefont {Valenti}\ \emph {et~al.}(2022)\citenamefont
  {Valenti}, \citenamefont {Jin}, \citenamefont {L\'eonard}, \citenamefont
  {Huber},\ and\ \citenamefont {Greplova}}]{PhysRevA.105.023302}%
  \BibitemOpen
  \bibfield  {author} {\bibinfo {author} {\bibfnamefont {A.}~\bibnamefont
  {Valenti}}, \bibinfo {author} {\bibfnamefont {G.}~\bibnamefont {Jin}},
  \bibinfo {author} {\bibfnamefont {J.}~\bibnamefont {L\'eonard}}, \bibinfo
  {author} {\bibfnamefont {S.~D.}\ \bibnamefont {Huber}},\ and\ \bibinfo
  {author} {\bibfnamefont {E.}~\bibnamefont {Greplova}},\ }\bibfield  {title}
  {\bibinfo {title} {Scalable hamiltonian learning for large-scale
  out-of-equilibrium quantum dynamics},\ }\href
  {https://doi.org/10.1103/PhysRevA.105.023302} {\bibfield  {journal} {\bibinfo
   {journal} {Phys. Rev. A}\ }\textbf {\bibinfo {volume} {105}},\ \bibinfo
  {pages} {023302} (\bibinfo {year} {2022})}\BibitemShut {NoStop}%
\bibitem [{\citenamefont {Che}\ \emph {et~al.}(2021)\citenamefont {Che},
  \citenamefont {Wei}, \citenamefont {Huang}, \citenamefont {Zhao},
  \citenamefont {Xue}, \citenamefont {Nie}, \citenamefont {Li}, \citenamefont
  {Lu},\ and\ \citenamefont {Xin}}]{PhysRevResearch.3.023246}%
  \BibitemOpen
  \bibfield  {author} {\bibinfo {author} {\bibfnamefont {L.}~\bibnamefont
  {Che}}, \bibinfo {author} {\bibfnamefont {C.}~\bibnamefont {Wei}}, \bibinfo
  {author} {\bibfnamefont {Y.}~\bibnamefont {Huang}}, \bibinfo {author}
  {\bibfnamefont {D.}~\bibnamefont {Zhao}}, \bibinfo {author} {\bibfnamefont
  {S.}~\bibnamefont {Xue}}, \bibinfo {author} {\bibfnamefont {X.}~\bibnamefont
  {Nie}}, \bibinfo {author} {\bibfnamefont {J.}~\bibnamefont {Li}}, \bibinfo
  {author} {\bibfnamefont {D.}~\bibnamefont {Lu}},\ and\ \bibinfo {author}
  {\bibfnamefont {T.}~\bibnamefont {Xin}},\ }\bibfield  {title} {\bibinfo
  {title} {Learning quantum hamiltonians from single-qubit measurements},\
  }\href {https://doi.org/10.1103/PhysRevResearch.3.023246} {\bibfield
  {journal} {\bibinfo  {journal} {Phys. Rev. Res.}\ }\textbf {\bibinfo {volume}
  {3}},\ \bibinfo {pages} {023246} (\bibinfo {year} {2021})}\BibitemShut
  {NoStop}%
\bibitem [{\citenamefont {Valenti}\ \emph {et~al.}(2019)\citenamefont
  {Valenti}, \citenamefont {van Nieuwenburg}, \citenamefont {Huber},\ and\
  \citenamefont {Greplova}}]{PhysRevResearch.1.033092}%
  \BibitemOpen
  \bibfield  {author} {\bibinfo {author} {\bibfnamefont {A.}~\bibnamefont
  {Valenti}}, \bibinfo {author} {\bibfnamefont {E.}~\bibnamefont {van
  Nieuwenburg}}, \bibinfo {author} {\bibfnamefont {S.}~\bibnamefont {Huber}},\
  and\ \bibinfo {author} {\bibfnamefont {E.}~\bibnamefont {Greplova}},\
  }\bibfield  {title} {\bibinfo {title} {Hamiltonian learning for quantum error
  correction},\ }\href {https://doi.org/10.1103/PhysRevResearch.1.033092}
  {\bibfield  {journal} {\bibinfo  {journal} {Phys. Rev. Res.}\ }\textbf
  {\bibinfo {volume} {1}},\ \bibinfo {pages} {033092} (\bibinfo {year}
  {2019})}\BibitemShut {NoStop}%
\bibitem [{\citenamefont {Koch}\ and\ \citenamefont
  {Lado}(2022)}]{PhysRevResearch.4.033223}%
  \BibitemOpen
  \bibfield  {author} {\bibinfo {author} {\bibfnamefont {R.}~\bibnamefont
  {Koch}}\ and\ \bibinfo {author} {\bibfnamefont {J.~L.}\ \bibnamefont
  {Lado}},\ }\bibfield  {title} {\bibinfo {title} {Designing quantum many-body
  matter with conditional generative adversarial networks},\ }\href
  {https://doi.org/10.1103/PhysRevResearch.4.033223} {\bibfield  {journal}
  {\bibinfo  {journal} {Phys. Rev. Res.}\ }\textbf {\bibinfo {volume} {4}},\
  \bibinfo {pages} {033223} (\bibinfo {year} {2022})}\BibitemShut {NoStop}%
\bibitem [{\citenamefont {Di~Franco}\ \emph {et~al.}(2009)\citenamefont
  {Di~Franco}, \citenamefont {Paternostro},\ and\ \citenamefont
  {Kim}}]{PhysRevLett.102.187203}%
  \BibitemOpen
  \bibfield  {author} {\bibinfo {author} {\bibfnamefont {C.}~\bibnamefont
  {Di~Franco}}, \bibinfo {author} {\bibfnamefont {M.}~\bibnamefont
  {Paternostro}},\ and\ \bibinfo {author} {\bibfnamefont {M.~S.}\ \bibnamefont
  {Kim}},\ }\bibfield  {title} {\bibinfo {title} {Hamiltonian tomography in an
  access-limited setting without state initialization},\ }\href
  {https://doi.org/10.1103/PhysRevLett.102.187203} {\bibfield  {journal}
  {\bibinfo  {journal} {Phys. Rev. Lett.}\ }\textbf {\bibinfo {volume} {102}},\
  \bibinfo {pages} {187203} (\bibinfo {year} {2009})}\BibitemShut {NoStop}%
\bibitem [{\citenamefont {Koch}\ \emph {et~al.}(2023)\citenamefont {Koch},
  \citenamefont {van Driel}, \citenamefont {Bordin}, \citenamefont {Lado},\
  and\ \citenamefont {Greplova}}]{PhysRevApplied.20.044081}%
  \BibitemOpen
  \bibfield  {author} {\bibinfo {author} {\bibfnamefont {R.}~\bibnamefont
  {Koch}}, \bibinfo {author} {\bibfnamefont {D.}~\bibnamefont {van Driel}},
  \bibinfo {author} {\bibfnamefont {A.}~\bibnamefont {Bordin}}, \bibinfo
  {author} {\bibfnamefont {J.~L.}\ \bibnamefont {Lado}},\ and\ \bibinfo
  {author} {\bibfnamefont {E.}~\bibnamefont {Greplova}},\ }\bibfield  {title}
  {\bibinfo {title} {Adversarial hamiltonian learning of quantum dots in a
  minimal kitaev chain},\ }\href
  {https://doi.org/10.1103/PhysRevApplied.20.044081} {\bibfield  {journal}
  {\bibinfo  {journal} {Phys. Rev. Appl.}\ }\textbf {\bibinfo {volume} {20}},\
  \bibinfo {pages} {044081} (\bibinfo {year} {2023})}\BibitemShut {NoStop}%
\bibitem [{\citenamefont {Schmale}\ \emph {et~al.}(2022)\citenamefont
  {Schmale}, \citenamefont {Reh},\ and\ \citenamefont
  {G\"{a}rttner}}]{Schmale2022}%
  \BibitemOpen
  \bibfield  {author} {\bibinfo {author} {\bibfnamefont {T.}~\bibnamefont
  {Schmale}}, \bibinfo {author} {\bibfnamefont {M.}~\bibnamefont {Reh}},\ and\
  \bibinfo {author} {\bibfnamefont {M.}~\bibnamefont {G\"{a}rttner}},\
  }\bibfield  {title} {\bibinfo {title} {Efficient quantum state tomography
  with convolutional neural networks},\ }\bibfield  {journal} {\bibinfo
  {journal} {npj Quantum Information}\ }\textbf {\bibinfo {volume} {8}},\ \href
  {https://doi.org/10.1038/s41534-022-00621-4} {10.1038/s41534-022-00621-4}
  (\bibinfo {year} {2022})\BibitemShut {NoStop}%
\bibitem [{\citenamefont {Neugebauer}\ \emph {et~al.}(2020)\citenamefont
  {Neugebauer}, \citenamefont {Fischer}, \citenamefont {J\"ager}, \citenamefont
  {Czischek}, \citenamefont {Jochim}, \citenamefont {Weidem\"uller},\ and\
  \citenamefont {G\"arttner}}]{PhysRevA.102.042604}%
  \BibitemOpen
  \bibfield  {author} {\bibinfo {author} {\bibfnamefont {M.}~\bibnamefont
  {Neugebauer}}, \bibinfo {author} {\bibfnamefont {L.}~\bibnamefont {Fischer}},
  \bibinfo {author} {\bibfnamefont {A.}~\bibnamefont {J\"ager}}, \bibinfo
  {author} {\bibfnamefont {S.}~\bibnamefont {Czischek}}, \bibinfo {author}
  {\bibfnamefont {S.}~\bibnamefont {Jochim}}, \bibinfo {author} {\bibfnamefont
  {M.}~\bibnamefont {Weidem\"uller}},\ and\ \bibinfo {author} {\bibfnamefont
  {M.}~\bibnamefont {G\"arttner}},\ }\bibfield  {title} {\bibinfo {title}
  {Neural-network quantum state tomography in a two-qubit experiment},\ }\href
  {https://doi.org/10.1103/PhysRevA.102.042604} {\bibfield  {journal} {\bibinfo
   {journal} {Phys. Rev. A}\ }\textbf {\bibinfo {volume} {102}},\ \bibinfo
  {pages} {042604} (\bibinfo {year} {2020})}\BibitemShut {NoStop}%
\bibitem [{\citenamefont {Torlai}\ \emph {et~al.}(2018)\citenamefont {Torlai},
  \citenamefont {Mazzola}, \citenamefont {Carrasquilla}, \citenamefont
  {Troyer}, \citenamefont {Melko},\ and\ \citenamefont {Carleo}}]{Torlai2018}%
  \BibitemOpen
  \bibfield  {author} {\bibinfo {author} {\bibfnamefont {G.}~\bibnamefont
  {Torlai}}, \bibinfo {author} {\bibfnamefont {G.}~\bibnamefont {Mazzola}},
  \bibinfo {author} {\bibfnamefont {J.}~\bibnamefont {Carrasquilla}}, \bibinfo
  {author} {\bibfnamefont {M.}~\bibnamefont {Troyer}}, \bibinfo {author}
  {\bibfnamefont {R.}~\bibnamefont {Melko}},\ and\ \bibinfo {author}
  {\bibfnamefont {G.}~\bibnamefont {Carleo}},\ }\bibfield  {title} {\bibinfo
  {title} {Neural-network quantum state tomography},\ }\href
  {https://doi.org/10.1038/s41567-018-0048-5} {\bibfield  {journal} {\bibinfo
  {journal} {Nature Physics}\ }\textbf {\bibinfo {volume} {14}},\ \bibinfo
  {pages} {447} (\bibinfo {year} {2018})}\BibitemShut {NoStop}%
\bibitem [{\citenamefont {Quek}\ \emph {et~al.}(2021)\citenamefont {Quek},
  \citenamefont {Fort},\ and\ \citenamefont {Ng}}]{Quek2021}%
  \BibitemOpen
  \bibfield  {author} {\bibinfo {author} {\bibfnamefont {Y.}~\bibnamefont
  {Quek}}, \bibinfo {author} {\bibfnamefont {S.}~\bibnamefont {Fort}},\ and\
  \bibinfo {author} {\bibfnamefont {H.~K.}\ \bibnamefont {Ng}},\ }\bibfield
  {title} {\bibinfo {title} {Adaptive quantum state tomography with neural
  networks},\ }\bibfield  {journal} {\bibinfo  {journal} {npj Quantum
  Information}\ }\textbf {\bibinfo {volume} {7}},\ \href
  {https://doi.org/10.1038/s41534-021-00436-9} {10.1038/s41534-021-00436-9}
  (\bibinfo {year} {2021})\BibitemShut {NoStop}%
\bibitem [{\citenamefont {Palmieri}\ \emph {et~al.}(2020)\citenamefont
  {Palmieri}, \citenamefont {Kovlakov}, \citenamefont {Bianchi}, \citenamefont
  {Yudin}, \citenamefont {Straupe}, \citenamefont {Biamonte},\ and\
  \citenamefont {Kulik}}]{Palmieri2020}%
  \BibitemOpen
  \bibfield  {author} {\bibinfo {author} {\bibfnamefont {A.~M.}\ \bibnamefont
  {Palmieri}}, \bibinfo {author} {\bibfnamefont {E.}~\bibnamefont {Kovlakov}},
  \bibinfo {author} {\bibfnamefont {F.}~\bibnamefont {Bianchi}}, \bibinfo
  {author} {\bibfnamefont {D.}~\bibnamefont {Yudin}}, \bibinfo {author}
  {\bibfnamefont {S.}~\bibnamefont {Straupe}}, \bibinfo {author} {\bibfnamefont
  {J.~D.}\ \bibnamefont {Biamonte}},\ and\ \bibinfo {author} {\bibfnamefont
  {S.}~\bibnamefont {Kulik}},\ }\bibfield  {title} {\bibinfo {title}
  {Experimental neural network enhanced quantum tomography},\ }\bibfield
  {journal} {\bibinfo  {journal} {npj Quantum Information}\ }\textbf {\bibinfo
  {volume} {6}},\ \href {https://doi.org/10.1038/s41534-020-0248-6}
  {10.1038/s41534-020-0248-6} (\bibinfo {year} {2020})\BibitemShut {NoStop}%
\bibitem [{\citenamefont {Ahmed}\ \emph {et~al.}(2021)\citenamefont {Ahmed},
  \citenamefont {S\'anchez Mu\~noz}, \citenamefont {Nori},\ and\ \citenamefont
  {Kockum}}]{PhysRevLett.127.140502}%
  \BibitemOpen
  \bibfield  {author} {\bibinfo {author} {\bibfnamefont {S.}~\bibnamefont
  {Ahmed}}, \bibinfo {author} {\bibfnamefont {C.}~\bibnamefont {S\'anchez
  Mu\~noz}}, \bibinfo {author} {\bibfnamefont {F.}~\bibnamefont {Nori}},\ and\
  \bibinfo {author} {\bibfnamefont {A.~F.}\ \bibnamefont {Kockum}},\ }\bibfield
   {title} {\bibinfo {title} {Quantum state tomography with conditional
  generative adversarial networks},\ }\href
  {https://doi.org/10.1103/PhysRevLett.127.140502} {\bibfield  {journal}
  {\bibinfo  {journal} {Phys. Rev. Lett.}\ }\textbf {\bibinfo {volume} {127}},\
  \bibinfo {pages} {140502} (\bibinfo {year} {2021})}\BibitemShut {NoStop}%
\bibitem [{\citenamefont {Koutn\'y}\ \emph {et~al.}(2022)\citenamefont
  {Koutn\'y}, \citenamefont {Motka}, \citenamefont {Hradil}, \citenamefont
  {\ifmmode \check{R}\else \v{R}\fi{}eh\'a\ifmmode~\check{c}\else
  \v{c}\fi{}ek},\ and\ \citenamefont {S\'anchez-Soto}}]{PhysRevA.106.012409}%
  \BibitemOpen
  \bibfield  {author} {\bibinfo {author} {\bibfnamefont {D.}~\bibnamefont
  {Koutn\'y}}, \bibinfo {author} {\bibfnamefont {L.}~\bibnamefont {Motka}},
  \bibinfo {author} {\bibfnamefont {Z.~c.~v.}\ \bibnamefont {Hradil}}, \bibinfo
  {author} {\bibfnamefont {J.}~\bibnamefont {\ifmmode \check{R}\else
  \v{R}\fi{}eh\'a\ifmmode~\check{c}\else \v{c}\fi{}ek}},\ and\ \bibinfo
  {author} {\bibfnamefont {L.~L.}\ \bibnamefont {S\'anchez-Soto}},\ }\bibfield
  {title} {\bibinfo {title} {Neural-network quantum state tomography},\ }\href
  {https://doi.org/10.1103/PhysRevA.106.012409} {\bibfield  {journal} {\bibinfo
   {journal} {Phys. Rev. A}\ }\textbf {\bibinfo {volume} {106}},\ \bibinfo
  {pages} {012409} (\bibinfo {year} {2022})}\BibitemShut {NoStop}%
\bibitem [{\citenamefont {Cha}\ \emph {et~al.}(2021)\citenamefont {Cha},
  \citenamefont {Ginsparg}, \citenamefont {Wu}, \citenamefont {Carrasquilla},
  \citenamefont {McMahon},\ and\ \citenamefont {Kim}}]{Cha2021}%
  \BibitemOpen
  \bibfield  {author} {\bibinfo {author} {\bibfnamefont {P.}~\bibnamefont
  {Cha}}, \bibinfo {author} {\bibfnamefont {P.}~\bibnamefont {Ginsparg}},
  \bibinfo {author} {\bibfnamefont {F.}~\bibnamefont {Wu}}, \bibinfo {author}
  {\bibfnamefont {J.}~\bibnamefont {Carrasquilla}}, \bibinfo {author}
  {\bibfnamefont {P.~L.}\ \bibnamefont {McMahon}},\ and\ \bibinfo {author}
  {\bibfnamefont {E.-A.}\ \bibnamefont {Kim}},\ }\bibfield  {title} {\bibinfo
  {title} {Attention-based quantum tomography},\ }\href
  {https://doi.org/10.1088/2632-2153/ac362b} {\bibfield  {journal} {\bibinfo
  {journal} {Machine Learning: Science and Technology}\ }\textbf {\bibinfo
  {volume} {3}},\ \bibinfo {pages} {01LT01} (\bibinfo {year}
  {2021})}\BibitemShut {NoStop}%
\bibitem [{\citenamefont {Wang}\ and\ \citenamefont {Kais}(2007)}]{Wang2007}%
  \BibitemOpen
  \bibfield  {author} {\bibinfo {author} {\bibfnamefont {H.}~\bibnamefont
  {Wang}}\ and\ \bibinfo {author} {\bibfnamefont {S.}~\bibnamefont {Kais}},\
  }\bibfield  {title} {\bibinfo {title} {Quantum entanglement and electron
  correlation in molecular systems},\ }\href
  {https://doi.org/10.1560/ijc.47.1.59} {\bibfield  {journal} {\bibinfo
  {journal} {Israel Journal of Chemistry}\ }\textbf {\bibinfo {volume} {47}},\
  \bibinfo {pages} {59} (\bibinfo {year} {2007})}\BibitemShut {NoStop}%
\bibitem [{\citenamefont {Tichy}\ \emph {et~al.}(2011)\citenamefont {Tichy},
  \citenamefont {Mintert},\ and\ \citenamefont {Buchleitner}}]{Tichy2011}%
  \BibitemOpen
  \bibfield  {author} {\bibinfo {author} {\bibfnamefont {M.~C.}\ \bibnamefont
  {Tichy}}, \bibinfo {author} {\bibfnamefont {F.}~\bibnamefont {Mintert}},\
  and\ \bibinfo {author} {\bibfnamefont {A.}~\bibnamefont {Buchleitner}},\
  }\bibfield  {title} {\bibinfo {title} {Essential entanglement for atomic and
  molecular physics},\ }\href {https://doi.org/10.1088/0953-4075/44/19/192001}
  {\bibfield  {journal} {\bibinfo  {journal} {Journal of Physics B: Atomic,
  Molecular and Optical Physics}\ }\textbf {\bibinfo {volume} {44}},\ \bibinfo
  {pages} {192001} (\bibinfo {year} {2011})}\BibitemShut {NoStop}%
\bibitem [{\citenamefont {Hofer}(2013)}]{Hofer2013}%
  \BibitemOpen
  \bibfield  {author} {\bibinfo {author} {\bibfnamefont {T.~S.}\ \bibnamefont
  {Hofer}},\ }\bibfield  {title} {\bibinfo {title} {On the basis set
  convergence of electron{\textendash}electron entanglement measures:
  helium-like systems},\ }\bibfield  {journal} {\bibinfo  {journal} {Frontiers
  in Chemistry}\ }\textbf {\bibinfo {volume} {1}},\ \href
  {https://doi.org/10.3389/fchem.2013.00024} {10.3389/fchem.2013.00024}
  (\bibinfo {year} {2013})\BibitemShut {NoStop}%
\bibitem [{\citenamefont {Esquivel}\ \emph {et~al.}(2015)\citenamefont
  {Esquivel}, \citenamefont {L{\'{o}}pez-Rosa},\ and\ \citenamefont
  {Dehesa}}]{Esquivel2015}%
  \BibitemOpen
  \bibfield  {author} {\bibinfo {author} {\bibfnamefont {R.~O.}\ \bibnamefont
  {Esquivel}}, \bibinfo {author} {\bibfnamefont {S.}~\bibnamefont
  {L{\'{o}}pez-Rosa}},\ and\ \bibinfo {author} {\bibfnamefont {J.~S.}\
  \bibnamefont {Dehesa}},\ }\bibfield  {title} {\bibinfo {title} {Correlation
  energy as a measure of non-locality: Quantum entanglement of helium-like
  systems},\ }\href {https://doi.org/10.1209/0295-5075/111/40009} {\bibfield
  {journal} {\bibinfo  {journal} {{EPL} (Europhysics Letters)}\ }\textbf
  {\bibinfo {volume} {111}},\ \bibinfo {pages} {40009} (\bibinfo {year}
  {2015})}\BibitemShut {NoStop}%
\bibitem [{\citenamefont {L\"{o}wdin}(1955)}]{Lwdin1955}%
  \BibitemOpen
  \bibfield  {author} {\bibinfo {author} {\bibfnamefont {P.-O.}\ \bibnamefont
  {L\"{o}wdin}},\ }\bibfield  {title} {\bibinfo {title} {Quantum theory of
  many-particle systems. i. physical interpretations by means of density
  matrices, natural spin-orbitals, and convergence problems in the method of
  configurational interaction},\ }\href
  {https://doi.org/10.1103/physrev.97.1474} {\bibfield  {journal} {\bibinfo
  {journal} {Physical Review}\ }\textbf {\bibinfo {volume} {97}},\ \bibinfo
  {pages} {1474} (\bibinfo {year} {1955})}\BibitemShut {NoStop}%
\bibitem [{\citenamefont {Siegbahn}\ \emph {et~al.}(1981)\citenamefont
  {Siegbahn}, \citenamefont {Alml\"{o}f}, \citenamefont {Heiberg},\ and\
  \citenamefont {Roos}}]{Siegbahn1981}%
  \BibitemOpen
  \bibfield  {author} {\bibinfo {author} {\bibfnamefont {P.~E.~M.}\
  \bibnamefont {Siegbahn}}, \bibinfo {author} {\bibfnamefont {J.}~\bibnamefont
  {Alml\"{o}f}}, \bibinfo {author} {\bibfnamefont {A.}~\bibnamefont
  {Heiberg}},\ and\ \bibinfo {author} {\bibfnamefont {B.~O.}\ \bibnamefont
  {Roos}},\ }\bibfield  {title} {\bibinfo {title} {The complete active space
  {SCF} ({CASSCF}) method in a newton{\textendash}raphson formulation with
  application to the {HNO} molecule},\ }\href
  {https://doi.org/10.1063/1.441359} {\bibfield  {journal} {\bibinfo  {journal}
  {The Journal of Chemical Physics}\ }\textbf {\bibinfo {volume} {74}},\
  \bibinfo {pages} {2384} (\bibinfo {year} {1981})}\BibitemShut {NoStop}%
\bibitem [{\citenamefont {Yang}\ and\ \citenamefont
  {Feiguin}(2017)}]{Yang2017}%
  \BibitemOpen
  \bibfield  {author} {\bibinfo {author} {\bibfnamefont {C.}~\bibnamefont
  {Yang}}\ and\ \bibinfo {author} {\bibfnamefont {A.~E.}\ \bibnamefont
  {Feiguin}},\ }\bibfield  {title} {\bibinfo {title} {Unveiling the internal
  entanglement structure of the kondo singlet},\ }\bibfield  {journal}
  {\bibinfo  {journal} {Physical Review B}\ }\textbf {\bibinfo {volume} {95}},\
  \href {https://doi.org/10.1103/physrevb.95.115106}
  {10.1103/physrevb.95.115106} (\bibinfo {year} {2017})\BibitemShut {NoStop}%
\bibitem [{\citenamefont {Zheng}\ \emph {et~al.}(2020)\citenamefont {Zheng},
  \citenamefont {He},\ and\ \citenamefont {Lu}}]{Zheng2020}%
  \BibitemOpen
  \bibfield  {author} {\bibinfo {author} {\bibfnamefont {R.}~\bibnamefont
  {Zheng}}, \bibinfo {author} {\bibfnamefont {R.}~\bibnamefont {He}},\ and\
  \bibinfo {author} {\bibfnamefont {Z.}~\bibnamefont {Lu}},\ }\bibfield
  {title} {\bibinfo {title} {Natural orbitals renormalization group approach to
  a kondo singlet},\ }\bibfield  {journal} {\bibinfo  {journal} {Science China
  Physics, Mechanics \& Astronomy}\ }\textbf {\bibinfo {volume} {63}},\ \href
  {https://doi.org/10.1007/s11433-019-1520-3} {10.1007/s11433-019-1520-3}
  (\bibinfo {year} {2020})\BibitemShut {NoStop}%
\bibitem [{\citenamefont {Debertolis}\ \emph {et~al.}(2021)\citenamefont
  {Debertolis}, \citenamefont {Florens},\ and\ \citenamefont
  {Snyman}}]{Debertolis2021}%
  \BibitemOpen
  \bibfield  {author} {\bibinfo {author} {\bibfnamefont {M.}~\bibnamefont
  {Debertolis}}, \bibinfo {author} {\bibfnamefont {S.}~\bibnamefont
  {Florens}},\ and\ \bibinfo {author} {\bibfnamefont {I.}~\bibnamefont
  {Snyman}},\ }\bibfield  {title} {\bibinfo {title} {Few-body nature of kondo
  correlated ground states},\ }\bibfield  {journal} {\bibinfo  {journal}
  {Physical Review B}\ }\textbf {\bibinfo {volume} {103}},\ \href
  {https://doi.org/10.1103/physrevb.103.235166} {10.1103/physrevb.103.235166}
  (\bibinfo {year} {2021})\BibitemShut {NoStop}%
\bibitem [{\citenamefont {White}(1992)}]{PhysRevLett.69.2863}%
  \BibitemOpen
  \bibfield  {author} {\bibinfo {author} {\bibfnamefont {S.~R.}\ \bibnamefont
  {White}},\ }\bibfield  {title} {\bibinfo {title} {Density matrix formulation
  for quantum renormalization groups},\ }\href
  {https://doi.org/10.1103/PhysRevLett.69.2863} {\bibfield  {journal} {\bibinfo
   {journal} {Phys. Rev. Lett.}\ }\textbf {\bibinfo {volume} {69}},\ \bibinfo
  {pages} {2863} (\bibinfo {year} {1992})}\BibitemShut {NoStop}%
\bibitem [{ITe()}]{ITensorurl}%
  \BibitemOpen
  \href@noop {} {\bibinfo  {journal} {\mbox{ITensor Library}
  http://itensor.org}\ }\BibitemShut {NoStop}%
\bibitem [{\citenamefont {Fishman}\ \emph {et~al.}(2022)\citenamefont
  {Fishman}, \citenamefont {White},\ and\ \citenamefont
  {Stoudenmire}}]{itensor}%
  \BibitemOpen
\bibfield  {journal} {  }\bibfield  {author} {\bibinfo {author} {\bibfnamefont
  {M.}~\bibnamefont {Fishman}}, \bibinfo {author} {\bibfnamefont {S.~R.}\
  \bibnamefont {White}},\ and\ \bibinfo {author} {\bibfnamefont {E.~M.}\
  \bibnamefont {Stoudenmire}},\ }\bibfield  {title} {\bibinfo {title} {{The
  ITensor Software Library for Tensor Network Calculations}},\ }\href
  {https://doi.org/10.21468/SciPostPhysCodeb.4} {\bibfield  {journal} {\bibinfo
   {journal} {SciPost Phys. Codebases}\ ,\ \bibinfo {pages} {4}} (\bibinfo
  {year} {2022})}\BibitemShut {NoStop}%
\bibitem [{DMR()}]{DMRGpyLibrary}%
  \BibitemOpen
  \href@noop {} {\bibinfo {title} {{DMRG}py library,
  {\url{https://github.com/joselado/dmrgpy}}}}\BibitemShut {NoStop}%
\bibitem [{zen()}]{zenodolink}%
  \BibitemOpen
  \href@noop {} {\bibinfo {title} {Code and data of the work available in
  {\url{https://doi.org/10.5281/zenodo.10090477}}}}\BibitemShut {NoStop}%
\bibitem [{app()}]{appendix}%
  \BibitemOpen
  \href@noop {} {\bibinfo {title} {See \hyperlink{Choice of local correlators
  and sample size}{Appendix} for details}}\BibitemShut {NoStop}%
\bibitem [{Note1()}]{Note1}%
  \BibitemOpen
  \bibinfo {note} {For edge sites the average is replaced for the single
  existing nth neighbor}\BibitemShut {NoStop}%
\bibitem [{\citenamefont {Jolliffe}\ and\ \citenamefont
  {Cadima}(2016)}]{Jolliffe2016}%
  \BibitemOpen
  \bibfield  {author} {\bibinfo {author} {\bibfnamefont {I.~T.}\ \bibnamefont
  {Jolliffe}}\ and\ \bibinfo {author} {\bibfnamefont {J.}~\bibnamefont
  {Cadima}},\ }\bibfield  {title} {\bibinfo {title} {Principal component
  analysis: a review and recent developments},\ }\href
  {https://doi.org/10.1098/rsta.2015.0202} {\bibfield  {journal} {\bibinfo
  {journal} {Philosophical Transactions of the Royal Society A: Mathematical,
  Physical and Engineering Sciences}\ }\textbf {\bibinfo {volume} {374}},\
  \bibinfo {pages} {20150202} (\bibinfo {year} {2016})}\BibitemShut {NoStop}%
\bibitem [{\citenamefont {Daimon}(2011)}]{Daimon2011}%
  \BibitemOpen
  \bibfield  {author} {\bibinfo {author} {\bibfnamefont {T.}~\bibnamefont
  {Daimon}},\ }\bibfield  {title} {\bibinfo {title} {Box{\textendash}cox
  transformation},\ }in\ \href {https://doi.org/10.1007/978-3-642-04898-2_152}
  {\emph {\bibinfo {booktitle} {International Encyclopedia of Statistical
  Science}}}\ (\bibinfo  {publisher} {Springer Berlin Heidelberg},\ \bibinfo
  {year} {2011})\ pp.\ \bibinfo {pages} {176--178}\BibitemShut {NoStop}%
\end{thebibliography}%

\end{document}